\documentclass[11pt,a4paper]{article}
\usepackage{jcappub}
\usepackage[utf8]{inputenc}
\usepackage{graphicx,amsfonts,color,comment,amsmath,hyperref,float}
\usepackage{outlines}
\usepackage{cleveref}
\usepackage{natbib}
\newcommand{\tauon}{$\tau$-lepton}
\newcommand{\tauons}{$\tau$-leptons}
\newcommand{\nupyprop}{\texttt{nuPyProp}}
\newcommand{\nuSpaceSim}{\texttt{nuSpaceSim}}
\usepackage{float}
\usepackage{bold-extra}
\usepackage{listings}
\usepackage{xcolor}
\usepackage{url}
\usepackage{lineno}

\definecolor{codegreen}{rgb}{0,0.6,0}
\definecolor{codegray}{rgb}{0.5,0.5,0.5}
\definecolor{codepurple}{rgb}{0.58,0,0.82}
\definecolor{backcolour}{rgb}{0.95,0.95,0.92}

\lstdefinestyle{mystyle}{
    backgroundcolor=\color{backcolour},   
    commentstyle=\color{codegreen},
    keywordstyle=\color{magenta},
    stringstyle=\color{codepurple},
    basicstyle=\ttfamily,
    breakatwhitespace=false,         
    breaklines=true,                 
    captionpos=b,                    
    keepspaces=true,                 
    numbers=none,                    
    numbersep=5pt,                  
    showspaces=false,                
    showstringspaces=false,
    showtabs=false,                  
    tabsize=2
}

\title{Neutrino propagation in the Earth and emerging charged leptons with \nupyprop}
\author[a]{Diksha Garg,}
\author[a]{Sameer Patel,}
\author[a]{Mary Hall Reno,}
\author[b]{Alexander Reustle,}
\author[c]{Yosui Akaike,}
\author[d]{Luis A. Anchordoqui,}
\author[e]{Douglas R. Bergman,}
\author[e]{Isaac Buckland,}
\author[f,g]{Austin L. Cummings}
\author[h]{Johannes Eser,}
\author[i,b]{Fred Garcia,}
\author[i,h]{Claire Gu\'epin,}
\author[f]{Tobias Heibges,}
\author[j]{Andrew Ludwig,}
\author[b]{John F. Krizmanic,}
\author[k]{Simon Mackovjak,}
\author[f]{Eric Mayotte,}
\author[f]{Sonja Mayotte,}
\author[h]{Angela V. Olinto,}
\author[d]{Thomas C. Paul}
\author[j]{Andr\'es Romero-Wolf,}
\author[f]{Fr\'ed\'eric Sarazin,}
\author[b]{Tonia M. Venters,}
\author[f]{Lawrence Wiencke,}
\author[g]{Stephanie Wissel}
\affiliation[a]{Department of Physics and Astronomy,
University of Iowa, Iowa City, IA 52242, USA}
\affiliation[b]{Laboratory for Astoparticle Physics, NASA/Goddard Space Flight Center, Greenbelt, MD 20771, USA}
\affiliation[c]{Waseda Research Institute for Science and Engineering, Waseda University, Tokyo 162-0044, Japan}
\affiliation[d]{Department of Physics and Astronomy, Lehman College, City University of New York, Bronx, NY 10468, USA}
\affiliation[e]{Department of Physics and Astronomy, University of Utah, Salt Lake City, UT 84112, USA}
\affiliation[f]{Department of Physics, Colorado School of Mines, Golden, CO 80401, USA}
\affiliation[g]{Department of Physics, Department of Astronomy and Astrophysics, Institute for Gravitation and the Cosmos, Pennsylvania State University, State College, PA 16801}
\affiliation[h]{Department of Astronomy and Astrophysics, University of Chicago, Chicago, IL 60637, USA}
\affiliation[i]{Department of Astronomy, University of Maryland, College Park, MD 20742, USA}
\affiliation[j]{Jet Propulsion Laboratory, Pasadena, CA 91109, USA}
\affiliation[k]{Department of Space Physics, Institute of Experimental Physics, Slovak Academy of Sciences, 040 01 Ko\v{s}ice, Slovakia}

\emailAdd{diksha-garg@uiowa.edu}

\abstract{Ultra-high-energy neutrinos serve as messengers of some of the highest energy astrophysical environments. Given that neutrinos are neutral and only interact via weak interactions, neutrinos can emerge from sources, traverse astronomical distances, and point back to their origins. Their weak interactions require large target volumes for neutrino detection. Using the Earth as a neutrino converter, terrestrial, sub-orbital, and satellite-based instruments are able to detect signals of neutrino-induced extensive air showers. In this paper, we describe the software code \texttt{nuPyProp} that simulates tau neutrino and muon neutrino interactions in the Earth and predicts the spectrum of the \tauons\ and muons that emerge. The \texttt{nuPyProp} outputs are lookup tables of charged lepton exit probabilities and energies that can be used directly or as inputs to the \texttt{nuSpaceSim} code designed to simulate optical and radio signals from extensive air showers induced by the emerging charged leptons. We describe the inputs to the code, demonstrate its flexibility and show selected results for \tauon\ and muon exit probabilities and energy distributions. The \texttt{nuPyProp} code is open source, available on github.}
\keywords{keywords}

\date{\today}

\begin{document}
\maketitle

\section{Introduction}
Most of what we know about the Universe is a result of studying photons traveling to the Earth from distant sources. From radio to gamma-rays, some of the most energetic phenomena in astrophysical processes deliver themselves in the form of electromagnetic radiation that can be detected from Earth. However, at higher photon energies ($E_\gamma \gtrsim 100$ GeV), the Universe starts to become opaque to photons as the number of photonic events decreases sharply.  This is because ultra-high-energy photons interact with the cosmic microwave background (CMB) and the extragalactic background light (EBL), producing $e^+  e^-$ pairs \cite{Brown:1973}. Ruffini et al. \cite{Ruffini:2015oha} calculate a rough limit of redshift $z\sim 0.03$ ($\sim 120$ Mpc) for ultra-high-energy (UHE) photon propagation ($E_\gamma>10^7$ GeV).

Another class of candidates with which to observe the deep Universe are protons and nuclei, however in transit to the Earth, these cosmic rays incur energy losses due to scattering and other interactions and due to deflections in magnetic fields by the virtue of their electric charge (see, e.g., ref. \cite{Anchordoqui:2018qom}). They also experience energy losses through interaction with the CMB. Neutrinos, on the other hand, are neutral and weakly interacting particles. Astrophysical processes that produce UHE cosmic rays should also produce neutrinos through production and decay of charged mesons \cite{Gaisser:1994yf,Learned:2000sw,Becker:2007sv,Anchordoqui:2013dnh}. Furthermore, cosmogenic neutrinos are produced when cosmic rays interact with cosmic photons in the CMB \cite{Berezinsky:1969erk,Stecker:1978ah,Anchordoqui:2007fi,Ahlers:2010fw,Kotera:2010yn,AlvesBatista:2018zui,Heinze:2019jou}. The resulting neutrino flux is the target of a number of current and future neutrino telescopes. Detection of the cosmogenic neutrino flux will help our understanding of the highest energy cosmic ray accelerators, source evolution and the composition of cosmic rays. This makes neutrinos important as a high energy astrophysical probe.

High energy protons interact with photons through photo-hadronic processes \cite{Mucke:2000}. They then produce pions, which eventually decay, producing neutrinos through the following chain of processes:
$    \pi^+ \to \mu^+ + \nu_\mu$; $\mu^+ \to \bar\nu_\mu + \nu_e + e^+$, and
    $\pi^- \to \mu^- + \bar\nu_\mu $; $  \mu^- \to \nu_\mu + \bar\nu_e + e^-$.
At production, the initial flavor ratios of neutrinos at the source is $N_e : N_\mu : N_\tau \sim 1:2:0$, where $N_i$ is $N_{\nu_i} + N_{\bar\nu_i}$. This means that the production of tau neutrinos at the source is highly suppressed. Through neutrino oscillations and flavor mixing over large cosmic distances, this ratio observed at Earth translates to $N_e : N_\mu : N_\tau \sim 1:1:1$ \cite{Learned:1994wg, Pakvasa:2007dc,Aartsen:2015, Bustamante:2019sdb,Song:2020nfh}. Hence, regardless of the source of production, the fluxes of neutrinos of all three flavors are expected to arrive at the Earth. Many neutrino observatories such as ANITA~\cite{ANITA:2020sds}, Baikal~\cite{Baikal:2020}, KM3Net~\cite{Km3net:2020bdg}, IceCube~\cite{IceCube:2021}, and the Pierre Auger Observatory~\cite{PierreAuger:2015ihf} collect data on astrophysical neutrinos. The design and development of new detectors such as GRAND \cite{Alvarez-Muniz:2018bhp}, BEACON \cite{Wissel:2020fav}, Trinity \cite{Otte:2019aaf}, TAMBO \cite{Romero-Wolf:2020pzh}, PUEO \cite{PUEO:2020bnn}, IceCube-Gen 2 \cite{IceCube-Gen2:2021} and POEMMA \cite{Olinto_2021} with increased sensitivities to neutrino fluxes can help determine even faint sources of astrophysical neutrino emitters (see ref. \cite{Ackermann:2022rqc} for an overview).

\begin{figure}
     \centering
     \includegraphics[width=0.45\textwidth]{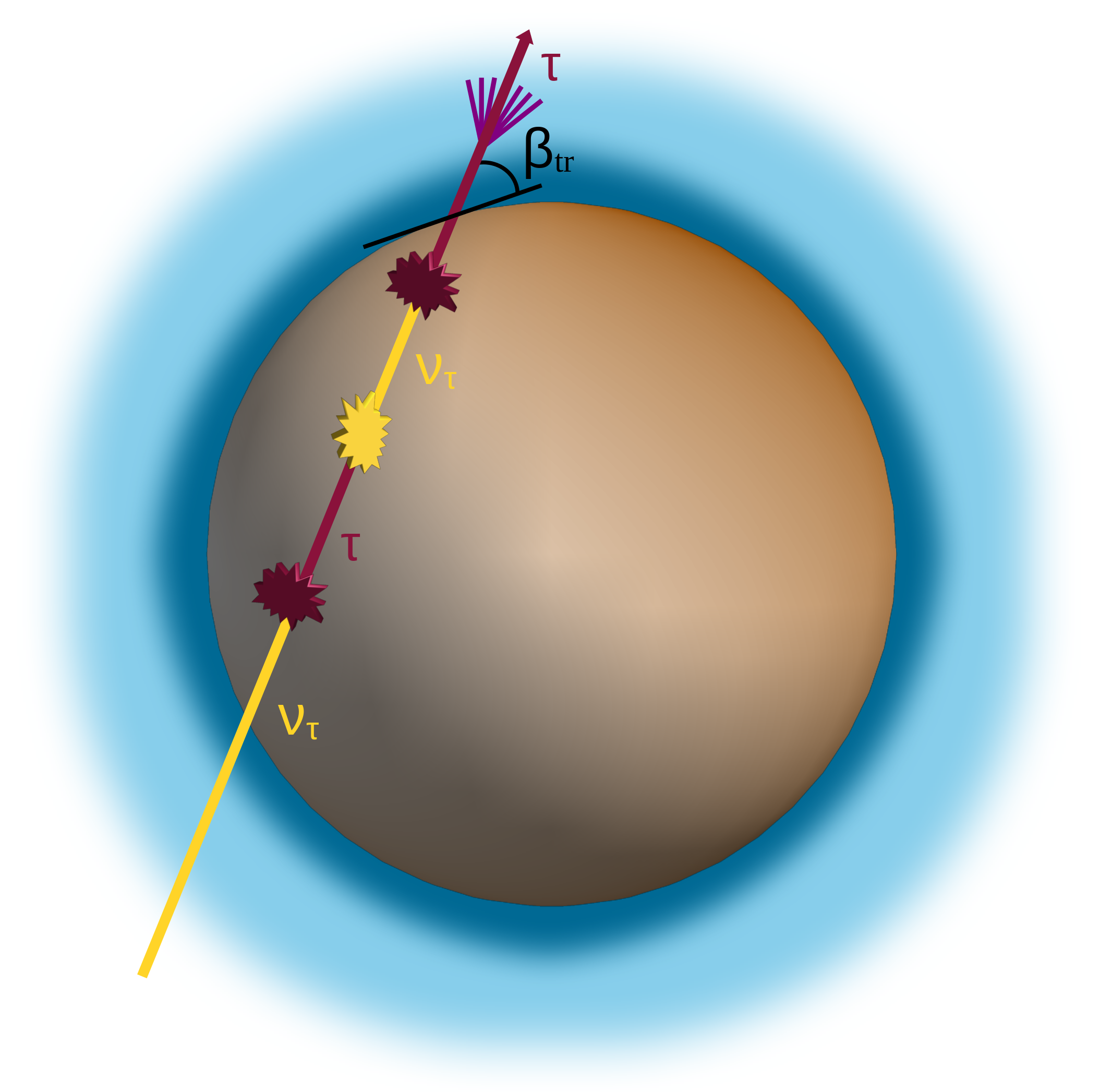}
     \includegraphics[width=0.45\textwidth]{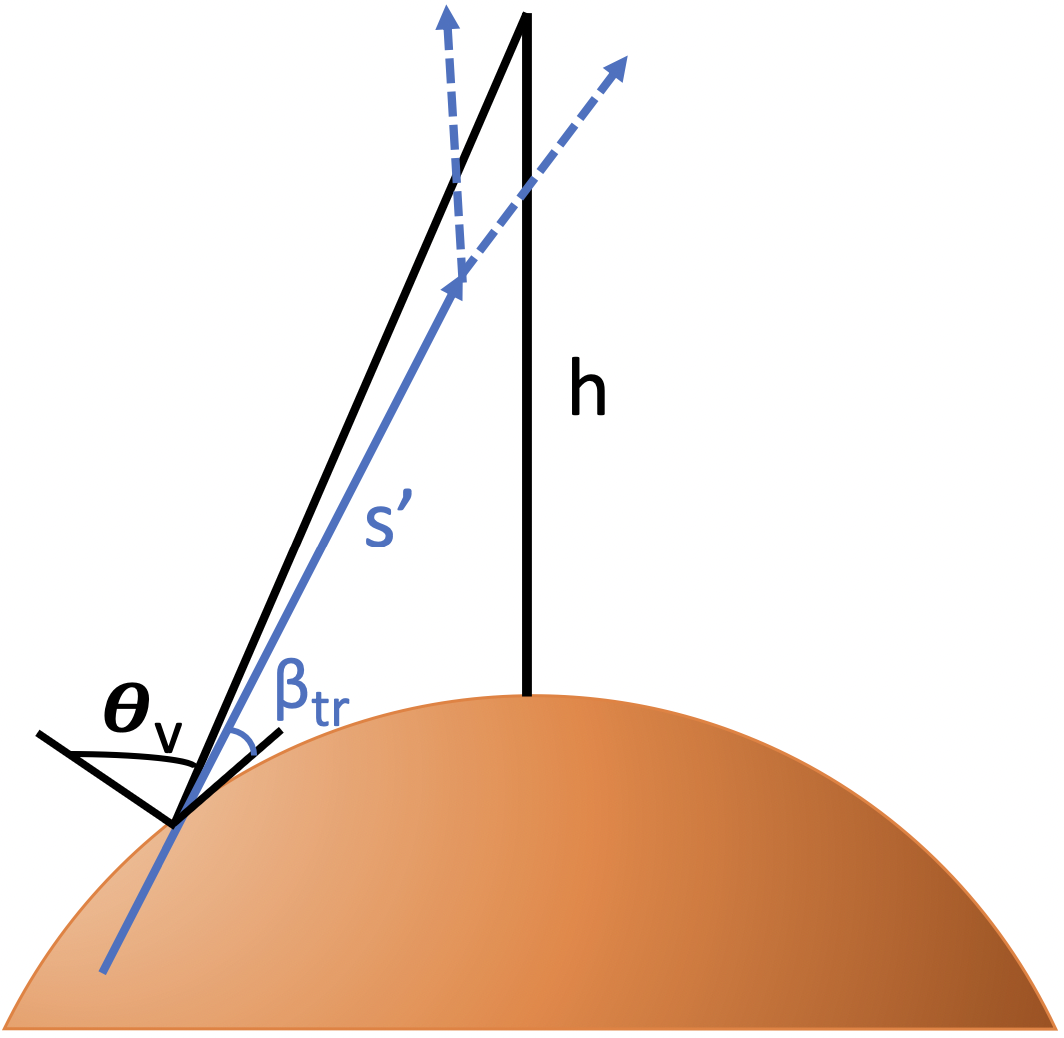}
     \caption{Schematic of incident tau neutrino trajectory that emerges as a {\tauon} at an angle $\beta_{tr}$ relative to a tangent to the Earth's surface. The depth of water is adjustable in the range of 0-10 km.}
     \label{fig:geometry}
     \vspace{-10pt}
 \end{figure}

Neutrino detection methods rely on a number of channels that depend on the incoming neutrino energy, angle and flavor. One particular method relies on Earth-emergent astrophysical tau neutrinos. Astrophysical tau neutrinos can be detected through upward-going extensive air showers (EASs) that are created in the Earth's atmosphere from \tauon\ decays \cite{Fargion:1999se,Fargion:2000iz,Fargion:2003kn,Fargion_2004}. These \tauons\ are produced by $\nu_\tau$ interactions inside the Earth while they transit through the Earth at relatively small slant angles, schematically shown in the left panel of \cref{fig:geometry}.  Because charged lepton electromagnetic energy losses scale as the inverse mass of the charged lepton, and because the \tauon\ lifetime is short, \tauons\ produced by $\nu_\tau$ interactions lose less energy in transit though the Earth than muons produced by $\nu_\mu$ interactions. The decays of \tauons\ lead to lower energy tau neutrinos (a process called $\nu_\tau$ regeneration) that can provide a substantial lower energy tau flux 
\cite{Halzen:1998be,Iyer:1999wu,Dutta:2000jv,Becattini:2000fj,Bugaev:2003sw,Bigas:2008sw}. 
Muons can also emerge from the Earth from muon neutrino interactions. While  regeneration is not a feature of $\nu_\mu$ propagation, muons also induce upward-going EASs \cite{Cummings:2020ycz}.

Space-based missions such as POEMMA \cite{POEMMA,Olinto:2020oky} and the sub-orbital instrument EUSO-SPB2 \cite{SPB2,Eser:2021mbp} can leverage the large surface area of the Earth to detect the beamed optical Cherenkov signals from cosmic neutrinos \cite{Reno:2019jtr,Venters:2019xwi,Reno:2021veo,Reno:2021xos} and perform complimentary observations of UHE cosmic rays using fluorescence telescopes. For determining the neutrino flux sensitivities of such missions, an end-to-end package to simulate the cosmic neutrino induced EASs becomes a necessity. \nuSpaceSim\  \cite{2019ICRC...36..936K,Krizmanic:2021eyu} is such a software, which is designed to simulate radio and optical signals from EASs that are induced by these astrophysical tau neutrinos and muon neutrinos. Presented here is a stand-alone component of \nuSpaceSim\ called {\nupyprop}\cite{nupyprop_github}. It takes incident
tau neutrinos and muon neutrinos and propagates them through the Earth. We use the more complex case of tau neutrino and \tauon\ propagation to illustrate how \nupyprop\ is structured and how \nupyprop\ contributes to the \nuSpaceSim\ software package.

Following the propagation and energy loss of tau neutrinos and \tauons\ through the Earth as in \cref{fig:geometry}, one can express the exiting tau observation probability in terms of the \tauon\ exit probability P$_{\text{exit}}$, the \tauon\ decay probability p$_{\text{decay}}$ for an infinitesimal path length $ds'$ in the atmosphere, and the detection probability p$_{\text{det}}$ \cite{Motloch:2013kva,Alvarez-Muniz:2018owm,Reno:2019jtr}:
\begin{equation}
    \text{P}_{\text{obs}}(E_{\nu_\tau},\beta_{tr},\theta_{\text{v}}) = \int \text{P}_{\text{exit}} (E_{\tau}|E_{\nu_{\tau}},\beta_{tr}) \times \left[\int ds'~\text{p}_{\text{decay}}(s')~\text{p}_{\text{det}}(E_{{\tau}},\theta_{\text{v}},\beta_{tr},s')\right] dE_{\tau},
    \label{eq:pobs}
\end{equation}
where $\beta_{tr}$ denotes the Earth emergence angle
(relative to tangent to the Earth at exit) of the $\tau$-lepton as shown in the right panel of \cref{fig:geometry} and $\theta_{\text{v}}$ is the angle along the line of sight from the point of Earth emergence to the detector. 
Here, p$_{\rm decay}$ relates to the decay of the \tauon\ in the Earth's atmosphere as a function of altitude (implicitly through $s'$, its path length in the atmosphere that depends on altitude and $\beta_{tr}$). The air shower produces light in the Cherenkov cone around the trajectory axis with an energy-weighted Cherenkov angle between $\sim 0.2^\circ-1.3^\circ$ depending on the altitude of the decay and $\beta_{tr}$ \cite{Reno:2019jtr}, so in general $\theta_{\rm v}\neq \beta_{tr}$. The quantity p$_{\rm det}$ accounts for how much of the shower signal is observed at the detector.

The \nuSpaceSim\ software package is designed to simulate \cref{eq:pobs} and determine, for example, the effective aperture for a given instrument. Modeling the \tauon\ P$_{\rm exit}$ is the first stage in the complete EAS simulation package. The \tauon\ exit probability is independent of the detector, so we have developed \nupyprop\ as a flexible, mission-independent simulation tool for P$_{\rm exit}(E_{\tau}|E_{\nu_{\tau}},\beta_{tr})$. 
The \nupyprop\  Monte-Carlo package generates lookup tables for exit probabilities and energy distributions for $\nu_\tau \to \tau$ and $\nu_\mu \to \mu$ propagation in the Earth. The average \tauon\ polarization is also generated. The look-up tables generated by \nupyprop\ are inputs to the \nuSpaceSim\ \cite{2019ICRC...36..936K} simulation of upward-going air showers. The \nupyprop\ package can, however, be installed and run independently. We provide instructions for installation and running in Appendix \ref{sec:install} and Appendix \ref{sec:usage}, and for customization of input lookup tables in Appendix \ref{app:custom}.

The \nupyprop\ software joins other codes to evaluate the propagation of neutrinos in the Earth, including {\texttt{TauRunner}} \cite{Safa:2021ghs}, \texttt{NuPropEarth} \cite{Garcia:2020jwr}, \texttt{PROPOSAL} \cite{Koehne:2013gpa,Dunsch:2018nsc}, \texttt{NuTauSim} \cite{Alvarez-Muniz:2018owm} and its update \texttt{NuLeptonSim} \cite{Cummings:NuLeptonSim}, and \texttt{Danton} \cite{Niess:2018opy}. Some comparisons of these codes can be found in ref. \cite{Abraham:2022jse}.
Features of the \nupyprop\ code are its modular construction that allow for user defined neutrino cross sections, charged lepton energy loss formulas and density modeling of the Earth. The focus is on stochastic energy loss for the charged lepton propagation, a feature that allows for an evaluation of the average polarization of the \tauons\ that emerge from the Earth which can be used in modeling the EASs from their decays. For comparison purposes, continuous energy loss for charged lepton propagation is also an option. As expected, for \tauon\ results for the exit probabilities are nearly identical for stochastic and continuous energy loss, however, the difference is $\sim 20\% $ for muons.

We begin in  \cref{sec:structure_framework} with an overview of the structure and framework of the \nupyprop\ package. \Cref{sec:earth} describes our inputs for the Earth density model. Neutrino and charged lepton interaction models are described and modeled in  \cref{sec:neutrino} and \cref{sec:lepton}. Regeneration and the treatment of tau decays is discussed in  \cref{sec:regeneration}. Finally, we show selected results in \cref{sec:results}, and conclude with a summary in \cref{sec:summary}. Supplementary details for our implementation of muon and \tauon\ propagation in the Earth are included in Appendix \ref{app:supplement}.

 \begin{figure}
     \centering
     \includegraphics[width=0.95\textwidth]{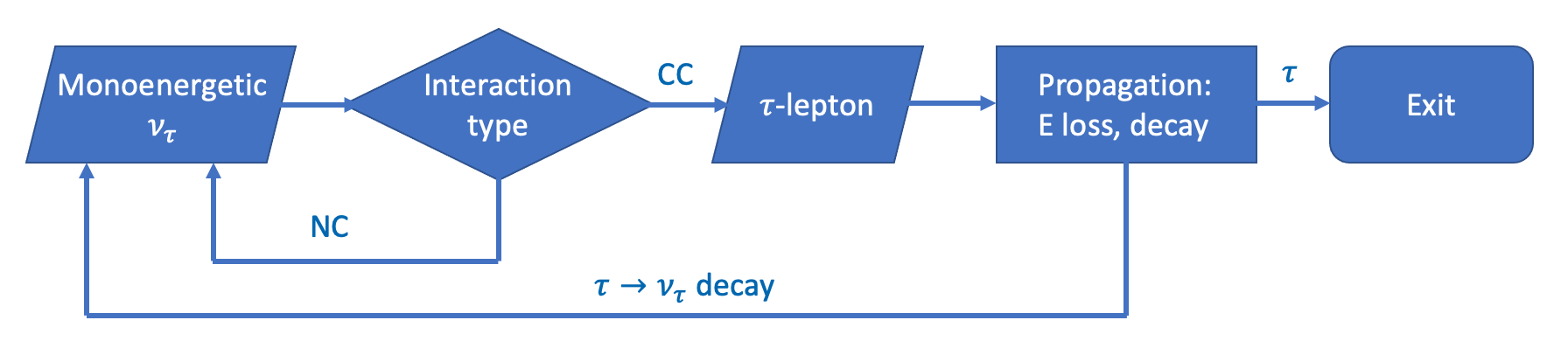}
     \caption{Flowchart of \nupyprop\ for tau neutrinos incident on the Earth which interact with charged-current (CC) and neutral-current (NC) interactions, to determine the \tauon\ exit probability and the energy with which it emerges. For $\nu_\mu$, the muon decay $\mu\to \nu_\mu e\bar\nu_{e}$ is not traced in the code since the decay $\nu_\mu$ have low energies.}
     \label{fig:flowchart}
     \vspace{-10pt}
 \end{figure}

\section{Structure and Framework of \nupyprop}
\label{sec:structure_framework}

The \nupyprop\ code has been designed to be flexible, with several neutrino and charged lepton interaction models and instructions to allow custom user inputs and models. Python is used for input and output procedures for lookup tables as well as for the creation of custom lookup tables for neutrino cross-sections and the charged lepton energy losses. The propagation part of the code is in Fortran and has been conveniently wrapped to Python using F2PY \cite{f2py}.  
The benefits of using Fortran for the propagation part of the package are as follows:
\begin{itemize}
    \item Pure Python (sometimes referred to as CPython) incorporates GIL \cite{gil}
    (Global Interpreter Lock) which prevents multiple threads from executing Python bytecodes at once. Since for the purposes of neutrino propagation, each incoming neutrino with a specific energy and angle can be propagated independently, it makes more sense that we leverage the use of multiple CPU cores (as per the computational resource availability) and cut the runtime of the code by running it parallely across multiple threads. No such limitation exists in Fortran and typically all inbuilt Fortran functions and datatypes are inherently threadsafe. Moreover, parallel computation in Fortran can be easily achieved by making use of OpenMP \cite{openmp}.
    \item One of the most popular modules for performing scientific and computational calculations in Python,  Numpy \cite{numpy} has been partly written in C and Fortran, making it ideal and fast for large scale computations. It shares a lot of common functionalities and paradigms with Fortran, thus provides a relatively easy and seamless integration with Fortran.
    \item F2PY is a part of Numpy and calling Fortran subroutines from Python is as easy and simple as calling regular Python functions after compilation of the Fortran code.
\end{itemize}

The core propagation flow that is followed by the code is shown in \cref{fig:flowchart}. We begin by injecting a number of monoenergetic tau neutrinos or muon neutrinos (or anti-neutrinos). These neutrinos first propagate through a layer of water and then into the subsequent inner layers of the Earth with material densities varying between those of rock and iron. Depending on the type of interaction, the neutrinos can either convert to same-flavor charged leptons through charged-current (CC) interactions or lose energy  through neutral-current  (NC) interaction with nucleons. 
The neutrino and antineutrino propagation through the Earth continues until a charged lepton is produced.

Through electromagnetic interactions, the charged leptons lose energy by ionization, bremsstrahlung, pair production and photonuclear processes
\cite{Lohmann:1985qg,Dutta:2000hh}. At high energies, we can take a one-dimensional approach for the charged particle trajectory \cite{Gutjahr:2022quk}. Charged leptons can also decay back into neutrinos with lower energies. This process is commonly known as regeneration. As muons are longer  lived than $\tau$-leptons, they lose a lot more energy than $\tau$-leptons before they decay. Due to this the regeneration process is only important for $\tau$-leptons. The \nupyprop\ program  tracks the charged leptons that make it through the Earth and records the energies of these outgoing particles. In addition, for each incident neutrino energy and angle, the exit probability $P_{\rm exit}$, final lepton energy, and average \tauon\ polarization upon exit are recorded. Although not yet included, future versions of the code will also track the exit probabilities and energies of neutrinos of all flavors.

\subsection{Lookup Tables}
\label{sec:lookup_tables}

Performing integrals at program runtime cost a great deal of computation time, especially in complex Monte Carlo propagation codes where these calculations have to be done more than a million times. The \nupyprop\ code output are lookup tables for \nuSpaceSim. 
For \nupyprop, to cut down the CPU time, input lookup tables for Earth column depths, neutrino and antineutrino cross sections and differential energy distributions, and energy loss cross sections and energy distributions are created prior to \nupyprop\ execution. While parameterization can be the fastest option, lookup tables are more flexible and straightforward to generate.
Template codes and default lookup tables are provided in the github code distribution. These lookup tables are interpolated at runtime. 
The provided lookup tables are neatly contained in a single Hierarchical Data Format (HDF), version 5 \cite{hdf5} file format.

\subsubsection{Input Lookup Tables}
\label{sec:input_lookup_tables}

The \nupyprop\ input lookup tables can be broadly divided into 3 categories:
\begin{enumerate}
    \item \textit{Earth trajectories} - These tables are used to calculate the column depth at a particular distance inside the Earth along the neutrino/charged lepton trajectory. Separate tables are provided for the water portion of the trajectory and the rest of the Earth. The water trajectory tables used are based on the user input water layer depth. Pre-built lookup tables for water layer depths of 0 km - 10 km (in units of 1 km) have been provided. Alternatively, users can generate their own water layer trajectories for use in the simulation. The details of Earth trajectories are discussed in  \cref{sec:earth}.
    \item \textit{Neutrino/anti-neutrino interactions} - These tables contain neutrino-nucleon and anti\-neutrino-nucleon cross sections for different parameterizations and parton model evaluations. The corresponding cumulative distribution functions (CDFs) for the inelasticity distributions for CC and NC interactions are provided. The inelasticity distributions are used for computing stochastic CC and NC interactions. The range of neutrino energies are from $10^3$ GeV to 10$^{12}$ GeV. For $E_\nu \gtrsim 10^3$ GeV, muon neutrino and tau neutrino cross sections are nearly identical and the neutrino cross sections with nucleons are well represented by evaluations using the parton model neglecting target mass effect \cite{Jeong:2010nt,Jeong:2010za,Feng:2022inv,Garcia:2020jwr}.  More discussion about the input neutrino cross sections and stochastic interactions is in \cref{sec:neutrino}.
    \item \textit{Charged lepton electromagnetic interactions} - These lookup tables account for electromagnetic energy loss of charged lepton, through average energy loss parameters used for $\langle dE/dX\rangle$ and cross sections and outgoing charged lepton energy distributions for stochastic energy losses
    \cite{Lohmann:1985qg,Dutta:2000hh}. 
    For all the tables, the charged lepton energy range is $10^3$ GeV$ < E < 10^{12}$ GeV. Stochastic and continuous energy losses for charged leptons are discussed in detail in  \cref{sec:stoch_losses} and \cref{sec:cont_losses}.
\end{enumerate}

\subsubsection{Output Lookup Tables}
\label{sec:output_lookup_tables}

All the output data from a single run of the code goes into a HDF output file. The output filename is a combination of the parameters set by the user at runtime and main contents of the output file can be categorized as:

\begin{enumerate}
    \item \textit{Exit probability} - This  set of tables contains the exit probabilities of the charged lepton as a function of the Earth emergence angles. It includes the probabilities with neutrino regeneration, and for reference, also includes  the charged lepton exit probabilities without neutrino regeneration.
    \item \textit{Charged lepton out-going energy CDFs} - These are tables of the CDFs of the out-going charged lepton energies calculated for the use of \nuSpaceSim, pre-binned for 71 log$_{10}$-bins in $z = E_{ch.lepton}/E_{\nu}$ for \tauons\ and 91 log$_{10}$-bins for muons. This range accounts for $E_\tau>10^5$ GeV and $E_\mu>10^3$ GeV in the exit CDFs. Options are provided in the code for users to define their own bins.   
    \item \textit{Charged lepton out-going energies} - For each incoming neutrino energy, optionally, sets of tables for each Earth emergence angle are written. The tables contain lists of each value of $\log_{10}{(E_{ch.lepton}/GeV)}$, where $E_{ch.lepton}$ is the energy of the charged lepton  that emerges from the Earth.
    \item \textit{Average polarization} - This set of table contains the average polarization of each exiting charged lepton as a function of the Earth emergence angles. It is only printed for stochastic energy loss and not continuous energy loss. The procedure to calculate average polarization is taken from \cite{Arguelles:2022bma}.  
\end{enumerate}

\subsection{Custom Models}
With an aim to make the simulation package more flexible, we provide an option for the user to add their own custom:
\begin{itemize}
    \item Column depth and water layer interpolation tables for user defined water layer depths.
    \item Neutrino (antineutrinos) cross-sections and inelasticity distributions in form of CDFs and cross section tables.
    \item Charged lepton cross-sections, inelasticity CDF distributions and the energy loss parameter for photonuclear energy loss based on different parameterizations of the electromagnetic structure function of nucleons.
\end{itemize}
Details of how to customize \nupyprop\ input lookup tables are included in Appendix \ref{app:custom}.

\section{Earth Model}
\label{sec:earth}

\begin{figure}
     \centering
     \includegraphics[width=0.45\textwidth]{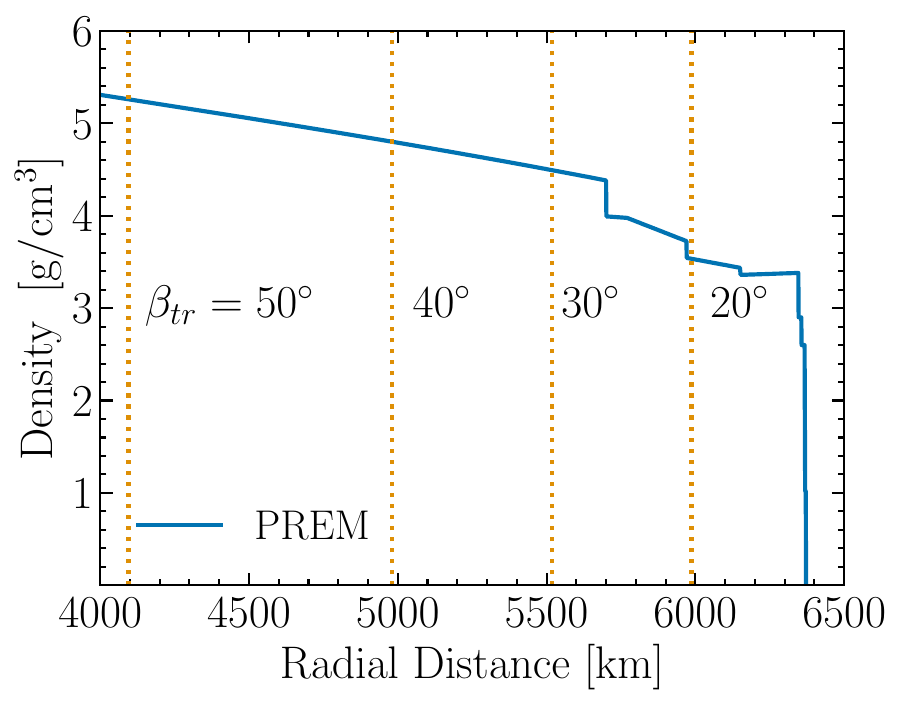}    
     \includegraphics[width=0.48\textwidth]{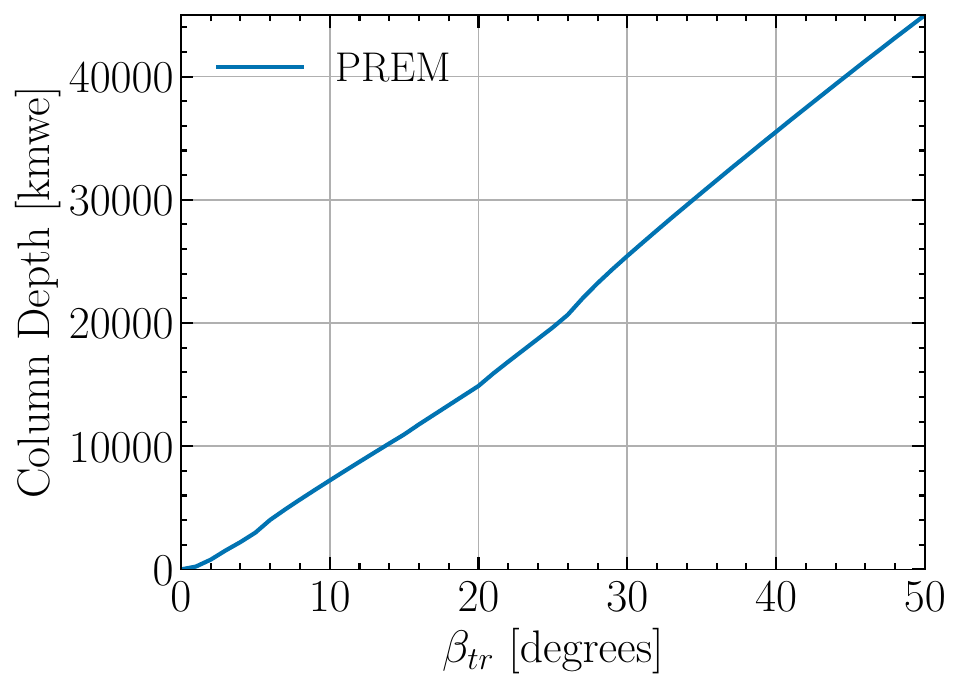}    
     \caption{Left: For the Preliminary Reference Earth Model (PREM) \cite{DZIEWONSKI1981297}, the Earth's density 
     as a function of radial distance from the center of the Earth. Vertical lines indicate the minimum radial distance of trajectories for several values of $\beta_{tr}$. Right: The column depth in units of kmwe = $10^5$ g/cm$^2$ as a function of Earth emergence angle $\beta_{tr}$.}
     \label{fig:prem}
     \vspace{-10pt}
 \end{figure}

The Earth's density is an important component in propagating charged particles inside the Earth because it  sets the amount of matter (i.e., target nucleons) that can interact with neutrinos and charged leptons. We use  the Preliminary Earth Reference Model (PREM) \cite{DZIEWONSKI1981297}.  The left panel of  \cref{fig:prem}  shows the PREM density of the Earth as a function of the radial distance from the center. The vertical lines indicated the radial distance of closest approach of trajectories with 
$\beta_{\rm tr}=20,$ 30, 40 and 50 degrees. For reference, the mantle-core boundary is at a radial distance $\sim 3500$ km, so the trajectories important here do not depend on details of the core density. The right panel of \cref{fig:prem} shows the total column depth $X_E$
\begin{equation}
    X_E(\beta_{\rm tr})=\int d\ell\, \rho\bigl(r(\ell,\beta_{\rm tr})\bigr)\,,
\end{equation}
from the density integrated along a neutrino/charged lepton trajectory $d\ell$, which depends  on the Earth emergence angle.

Keeping a fixed radius of a spherical Earth ($R_E=6371$ km), the depth of the water layer is a free parameter for the user to choose.
The provided lookup tables have been generated and included with the code package for depths from 0 km to 10 km in 1 km increments. The default PREM water depth is 4 km, as in, for example, simulations for the ANITA experiment \cite{Alvarez-Muniz:2017mpk}. Here, 0 km water layer means the charged leptons are exiting the Earth's surface without going through any water layers, meaning the Earth's crust extends to $R_E$. The results shown below are all performed with the water depth set to 4 km.

\section{Neutrino Interactions}
\label{sec:neutrino}

One of the key input parameter that accounts for some of the uncertainties in propagating neutrinos inside the Earth is the neutrino cross section. There are direct measurements of the neutrino and antineutrino cross sections below 400 GeV \cite{ParticleDataGroup:2020ssz}. In the range up to $\sim 10^6$ GeV, neutrino flux attenuation in the Earth and its impact on events in IceCube yield indirect measurements of the neutrino cross section \cite{IceCube:2017roe,Bustamante:2017xuy, IceCube:2020rnc} (see also \cite{Ackermann:2022rqc}).
In the neutrino energy range of $\sim 10^6-10^{12}$ GeV, the cross sections are not directly probed, however, energy and angular distributions of neutrino events in neutrino telescopes will help pin down standard model neutrino cross sections at high- and ultra-high energies \cite{Valera:2022ylt,Esteban:2022uuw,Ackermann:2022rqc}.  

The neutrino CC cross section for scattering with a nucleon N can be written in terms of Bjorken-$x$ defined $x\equiv Q^2/2p\cdot q$, the neutrino inelasticity $y\equiv p\cdot q/p\cdot k$  and structure functions $F_i(x,Q^2)$ for 
$\nu_\ell(k)+ N(p)\to \ell (k')+X(p_W)$ 
as \cite{Albright:1974ts,Kretzer:2002fr}
\begin{eqnarray} \nonumber
\frac{d^2\sigma^{\nu N}}{dx\ dy} &=& \frac{G_F^2 M
E_{\nu}}{\pi(1+Q^2/M_W^2)^2}
\Biggl(
(y^2 x + \frac{m_{\ell}^2 y}{2 E_{\nu} M})
F_1 + \left[ (1-\frac{m_{\ell}^2}{4 E_{\nu}^2})
-(1+\frac{M x}{2 E_{\nu}}) y\right]
F_2
\\ 
&+& 
\left[x y (1-\frac{y}{2})-\frac{m_{\ell}^2 y}{4 E_{\nu} M}\right]
F_3+
\frac{m_{\ell}^2(m_{\ell}^2+Q^2)}{4 E_{\nu}^2 M^2 x} F_4
- \frac{m_{\ell}^2}{E_{\nu} M} F_5
\Biggr)\, ,
\label{eq:nusig}
\end{eqnarray} 
with $Q^2\equiv -q^2=-(k-k')^2$, $(k')^2=m_\ell^2$ and $p^2=M^2$. The antineutrino CC cross section has the opposite sign of the term with $F_3$. The CC cross sections with tau neutrinos and anti-neutrinos will be slightly lower than for muon neutrinos and anti-neutrinos because of kinematic corrections to the range of integration and corrections of order $m_\tau^2/(M E_{\nu})$  in eq. (\ref{eq:nusig}). 
For $E_{\nu} = 10^3$ GeV, the tau neutrino and tau anti-neutrino CC cross-sections are 94.5\% and 92.6\% of the muon neutrino and muon anti-neutrino CC cross sections respectively \cite{Jeong:2010nt}. At $10^4$ GeV, the tau neutrino and tau anti-neutrino CC cross-sections are 98.5\% and 98.0\% of that of their muon counterparts respectively. We set $m_\ell = 0$ in our evaluation of the CC cross section for the input lookup tables that run from $E_\nu=10^3-10^{12}$ GeV. Since NC neutrino scattering has no mass dependence, the $\nu_\mu$ and $\nu_\tau$ NC cross sections are equal.

The usual approach to the evaluation of the structure functions in eq. (\ref{eq:nusig}) is to use the QCD improved parton model
\cite{Gandhi:1995tf,Gandhi:1998ri}. Treating the target nucleons as comprised of valence quarks, the quark and antiquark sea, and gluons, the structure functions depend on universal parton distribution functions (PDFs) that are extracted from a range of high energy physics data. The PDFs at different energy scales are related by the Dokshitzer-Gribov-Lipatov-Altarelli-Parisi (DGLAP) equations
(see, e.g., ref. \cite{Devenish:2004pb} for a review). A number of groups provide PDFs that largely agree in the kinematic region relevant to the neutrino cross section for energies up to $E_\nu\sim 10^6-10^7$ GeV. Their different extrapolations into kinematic regions not yet measured lead to a range of neutrino cross sections at higher energies. 

\begin{figure}
     \centering
     \includegraphics[width=0.48\textwidth]{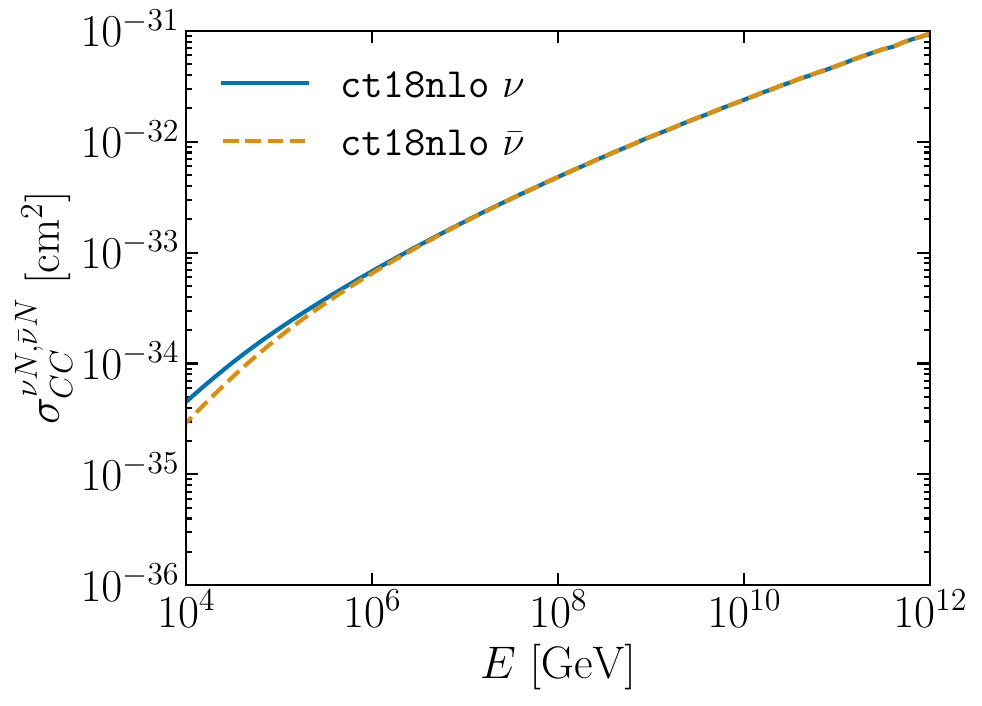}
     \includegraphics[width=0.48\textwidth]{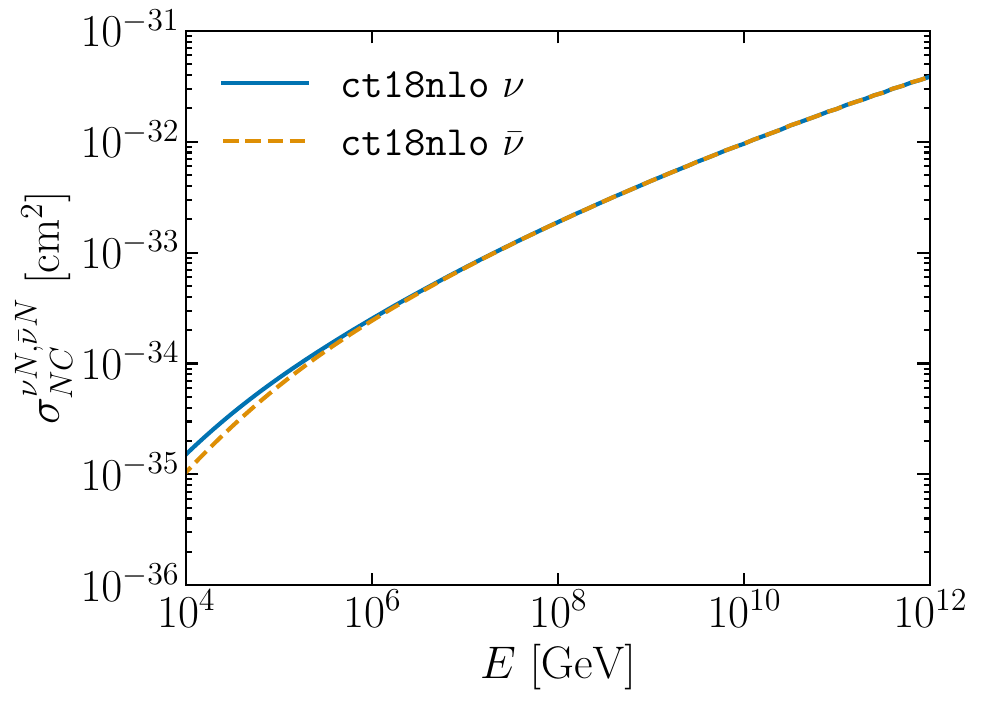}
     \caption{The neutrino (solid curves) and anti-neutrino (dashed curves) charged current (left) and neutral current (right) cross sections for interactions with isoscalar nucleons, as a function of energy using the CTEQ18-NLO PDFs \cite{Hou:2019efy} assuming isospin symmetry.}
     \label{fig:sigmas}
     \vspace{-10pt}
\end{figure}

In the lookup tables, the PDF-based neutrino cross sections are evaluated using PDFs that are next-to-leading order (NLO) in QCD  integrated with the leading-order cross section for neutrino/antineutrino scattering with quarks and antiquarks. Three PDF sets are considered here: CTEQ18-NLO \cite{Hou:2019efy} (\texttt{ct18nlo}), NCTEQ15 \cite{Kovarik:2015cma} (\texttt{nct15}) for free protons and
the MSTW \cite{Martin:2009iq} PDFs used by Connolly, Thorne and Waters (\texttt{ctw}) in ref. \cite{Connolly:2011vc}. At the high energies of interest here, the full NLO cross section evaluation for neutrino/antineutrino scattering and leading order evaluation using NLO PDFs differ by less than $\sim 5\%$ \cite{Jeong:2010za}. For CC interactions,  top quark production contributes  $\sim$ 10\% of the cross section at the highest energies \cite{Gluck:2010rw,Garcia:2020jwr}. At neutrino energies $E_{\nu} \gtrsim 10^7$ GeV, neutrinos and anti-neutrinos have the same cross sections, as shown in \cref{fig:sigmas} for charged current and neutral current interactions for \texttt{ct18nlo}. We refer to neutrinos and anti-neutrinos as ``neutrinos" henceforth. 

Another approach to the high energy behavior of the neutrino cross section is to rely on a formalism that encodes the asymptotic behavior of the high energy cross section to preserve unitarity \cite{Block:2013mia,Block:2013nia,Block:2014kza,Jeong:2014mla,Jeong:2017mzv,Arguelles:2015wba}. In a series of papers \cite{Block:2013mia,Block:2013nia,Block:2014kza}, Block et al. connected electromagnetic deep inelastic scattering, $\gamma p$ and neutrino inelastic scattering to parameterize the high energy neutrino cross section. The Block et al. neutrino cross sections are labeled \texttt{bdhm}
cross sections. 

Finally, a parameterization of the electromagnetic structure function $F_2$ determined from scattering data at HERA by Abramowicz et al. \cite{Abramowicz:1991xz,Abramowicz:1997ms} can be adapted to apply to the high energy neutrino cross section \cite{Jeong:2017mzv}. 
The electromagnetic and weak structure functions have the same small-$x$, large-$Q$ behavior up to an overall normalization. 
The neutrino cross section also depends on $F_i(x,Q)$ for $i=1,3-5$. Here, we use the Callan-Gross relation, $F_2(x,Q)=2xF_1(x,Q)$, to obtain $F_1(x,Q)$. The $F_3-F_5$ structure functions have negligible contributions at high energies. We find the overall
normalization by evaluating the neutrino cross section using the
electromagnetic \texttt{allm} $F_2$ and the Callan-Gross relation for $F_1$, then finding the normalization factor so that the neutrino cross section equals the \texttt{ct18nlo} neutrino cross section at $E_\nu=10^7$ GeV. 
This neutrino cross section is denoted as \texttt{allm}. A unified treatment of the high energy neutrino cross section and high energy charged lepton energy loss could employ, e.g., the \texttt{allm} structure functions in both evaluations. 

\begin{figure}
     \centering
     \includegraphics[width=0.48\textwidth]{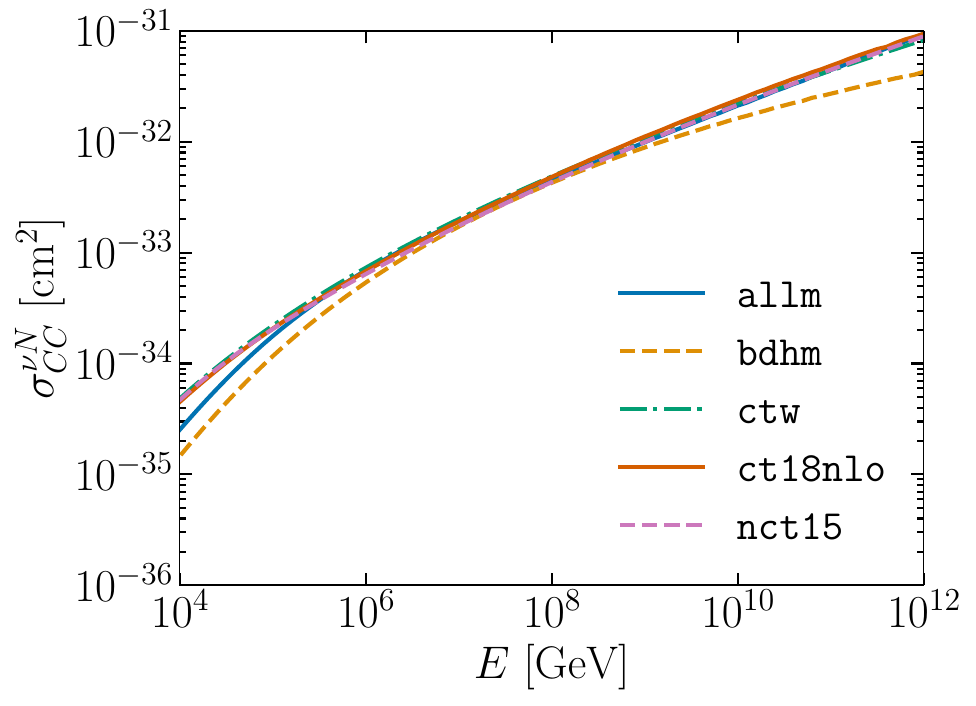}
     \includegraphics[width=0.48\textwidth]{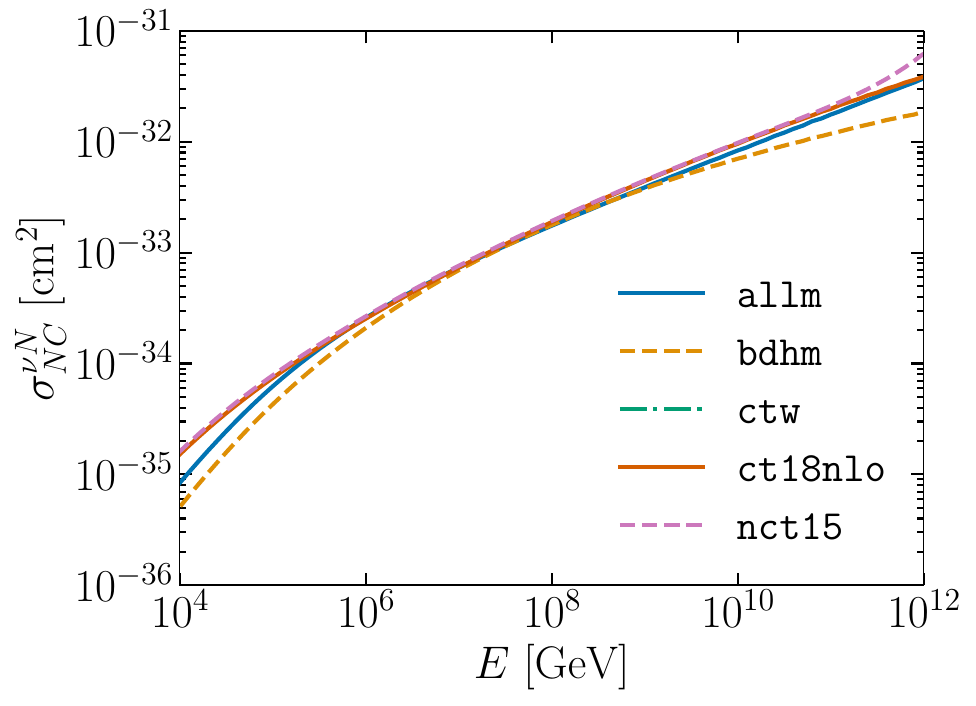}
     \caption{The neutrino charged current (left figure) and neutral current (right figure) cross sections for interactions with isoscalar nucleons, as a function of energy for the cross section models included in the \nupyprop\ distribution. See text for discussion of normalization of the \texttt{allm} cross sections.}
     \label{fig:sigmas2}
     \vspace{-10pt}
\end{figure}

\Cref{fig:sigmas2} shows a comparison of all the neutrino cross sections provided here. The cross sections  \texttt{ct18nlo} and \texttt{nct15} are nearly identical. The \texttt{bdhm} and \texttt{allm} cross sections do not have valence quark contributions included. This accounts for their lower cross sections at $E_\nu=10^4$ GeV. In addition, the neutrino and antineutrino cross sections are set equal. The \texttt{bdhm} and \texttt{allm} cross sections should not be used for $E_\nu\lesssim 10^6$ GeV. 

The neutrino or antineutrino interaction length (in units of g/cm$^2$) for targets with $A$ g/mole and $\sigma_{tot}=\sigma_{CC}+\sigma_{NC}$, is
\begin{equation}
\label{eq:int_depth_nu}
    X_{int} = (N_A \sigma_{tot}^{\nu A}/A)^{-1}\
\end{equation}
where $N_A = 6.022 \times 10^{23}$ particles per mole is Avogadro's number. Our evaluation of neutrino interactions in the Earth approximates
\begin{equation}
\label{eq:sigperA}
    \sigma_{tot}^{\nu A}/A \simeq \sigma_{tot}^{\nu N}
\end{equation}
for isoscalar nucleons $N$. At high energies, the neutrino cross section with protons equals the cross section with neutrons because the cross section is dominated by the quark and antiquark sea, not the valence quarks. 

There are, in principle, nuclear corrections associated with the fact that nucleons in nuclei are not free nucleons.  A complete understanding of nuclear corrections in neutrino scattering is not yet achieved, as discussed in, e.g., refs. \cite{Kopeliovich:2012kw,Muzakka:2022wey,Garcia:2020jwr} and references therein. Using the nuclear PDFs from the nCTEQ group \cite{Kovarik:2015cma}, the cross section per nucleon for neutrino scattering on aluminum ($A=27$, similar to silicon with $A=28$) is only a few percent lower than the cross section in eq. (\ref{eq:sigperA}) evaluated using isoscalar nucleons for $E_\nu\lesssim 10^5$ GeV. The neutrino cross section per nucleon scattering with aluminum is less than the cross section with isoscalar nucleons by $\sim 6\%$ for $E_\nu=10^6$ GeV, and the difference between the two changes by $3-4\%$ for each decade of energy. At $E_\nu=10^9$ GeV, the cross section per nucleon with aluminum is $\sim 17\%$ lower than with isoscalar nucleons using the nCTEQ15 aluminum and proton PDFs, respectively, and as much as $\sim 25\%$ lower for $E_\nu=10^{12}$ GeV. Similar or slightly smaller cross sections per nucleon are obtained for neutrino scattering with iron.
The current version of \nupyprop\ neutrino propagation in the Earth, described below, relies on the Earth's column depth and not on density dependent layers, using isoscalar nucleon targets. A $(Z,A)$ dependent neutrino cross section is not straightforward to implement in \nupyprop\ in its current form. Future studies of theoretical uncertainties for \nupyprop\ will account for the nuclear dependence of the cross section. It has been noted that future UHE neutrino detectors may be able to constrain the cross section at the level of these nuclear correction to UHE neutrino scattering \cite{Garcia:2020jwr,Valera:2022ylt,Esteban:2022uuw}.

Neutrinos and antineutrinos are propagated stochastically in \nupyprop. With each cross section, there is a corresponding CDF to generate the out-going lepton energy distribution according to the differential cross section for CC as in eq. (\ref{eq:nusig}) or for NC scattering.
Five pre-calculated sets of neutrino and antineutrino CC and NC cross-sections are part of the input lookup tables for {\nupyprop}. 
The CTW cross-sections have been calculated using the parameterization provided in \cref{eq:xc_nu} and the differential cross sections in ref. \cite{Connolly:2011vc}.
For reference, we also provide parameterizations of the neutrino and antineutrino cross sections as a function of energy for \texttt{allm}, \texttt{bdhm}, \texttt{ct18nlo} and \texttt{nct15}. External lookup tables are provided in \nupyprop\ for the CTW cross sections and CDFs along with a template (in \texttt{models.py}) for users to add custom neutrino cross sections. 

\subsection{The \texttt{propagate\_nu} subroutine}

The subroutine \texttt{propagate\_nu} resides in the FORTRAN code in {\nupyprop}. It propagates neutrinos stochastically along their trajectories in the Earth of column depth $X_E(\beta_{tr})$. Along $X_E$, using standard Monte Carlo techniques \cite{ParticleDataGroup:2020ssz}, the probability of interaction is generated according to an exponential distribution $\exp(-X/X_{int})$, the interaction type (CC or NC) is determined by their relative fractions of the $\sigma^{\nu N}_{tot}$ and the outgoing charged lepton (CC interaction) or neutrino (NC interaction) energy $E_f$ is determined according to the lookup table CDFs tabulated in terms of the inelasticity $y$ defined by
\begin{equation}
\label{eq:inelasticity}
    y = \frac{E_0 - E_f}{E_0}
\end{equation}
for interacting neutrino energy $E_0$ in the nucleon target rest frame. 
The \texttt{propagate\_nu} loop continues until one of the following three cases evaluates to True:
1) The total column depth is reached (which depends on the Earth emergence angle);
2)  The neutrino converts to a charged lepton through a CC interaction; or 
3) The neutrino energy reduces to a minimum value of $E_\nu = 10^3$ GeV.
The code follows the trajectory without accounting for the scattering angle in neutrino interactions. This angle starts to become important as the neutrino energy decreases. In $\nu_\mu$ CC interactions, the scattering angle for a muon produced with $E_\mu=10^3$ GeV from a neutrino with $E_\mu/(1-\langle y\rangle)$ is $\sim 0.5^\circ$ \cite{Gutjahr:2022quk}. For muons with energies of $\sim 10^3-10^4$, angular corrections may need to be applied to the \nupyprop\ results.

When the neutrino converts to a charged lepton, the \texttt{nuPyProp} code moves to charged lepton propagation in the Earth.

\section{Charged Lepton Propagation in the Earth}
\label{sec:lepton}

Charged current neutrino interactions yield their associated charged leptons. These charged leptons can decay, and prior to decay, they are subject to electromagnetic interactions: bremsstrahlung, electron-positron pair production and photonuclear interactions. 
At high energies, just as with the cross-section of neutrinos, the high energy extrapolations of the photonuclear cross section are uncertain. Hence, we provide lookup tables for multiple photonuclear energy loss models that account for different parameterizations of the electromagnetic structure function $F_2$. These different models are summarized in \cref{tab:nuceloss}.

\begin{table}[]
    \centering
    \begin{tabular}{|c|c|}
    \hline
    Photonuclear Energy Loss Model  & Reference\\
    \hline
    \hline
     Abramowicz, Levin, Levy, Maor (\texttt{allm})    & \cite{Abramowicz:1991xz,Abramowicz:1997ms} \\
     \hline
     Bezrukov, Bugaev (\texttt{bb})
         & \cite{Bezrukov:1981ci}\\
         \hline
     Block, Durand, Ha, McKay (\texttt{bdhm}) & \cite{Block:2014kza}\\
     \hline
     Capella, Kaidalov, Merino, Tran (\texttt{ckmt}) &
     \cite{Capella:1994cr}\\
     \hline
     User defined & \texttt{models.py}\\
     \hline
    \end{tabular}
    \caption{Photonuclear energy loss models \texttt{allm}, \texttt{bb}, \texttt{bdhm} and \texttt{ckmt} included in \nupyprop\ input lookup tables. }
    \label{tab:nuceloss}
\end{table}

A starting point is the average energy loss per column depth
of a charged lepton. It can be written as \cite{ParticleDataGroup:2020ssz,Dutta:2000hh}:
\begin{equation}
\label{eq:energy_loss}
    -\Bigg\langle\frac{dE}{dX}\Bigg\rangle = a ^\ell+ \sum\limits_{i={\rm brem, pair, nuc}} b_i^\ell(E) E\ ,
\end{equation}
where $X$ is the column depth travelled by the charged lepton $\ell$ with initial energy E,  
$a^\ell$ is the ionization energy loss, and
$b^\ell_i$ denotes energy loss from bremmstrahlung, pair production and photonuclear processes.
The energy loss parameter $b^\ell_i$ is dependent on the charged lepton energy and is given by:
\begin{equation}
\label{eq:beta}
    b_i^\ell(E) = \frac{N_A}{A} \int_{y_{min}}^{y_{max}} y\ \frac{d\sigma_i^{\ell N}(y,E)}{dy}\ \mathrm{d}y\,.
\end{equation}
Here, $y$ is the inelasticity parameter as in eq. \ref{eq:inelasticity} with $E_0$ equal to the initial charged lepton energy (E), and
$A$ is the atomic mass number of the target nucleus.

A summary of formulas for evaluating $a$ and $b_i$ are in refs. \cite{Lohmann:1985qg} and \cite{Dutta:2000hh}. An important uncertainty in average electromagnetic energy loss of \tauons\ and muons is the size and energy dependence of the photonuclear process \cite{Jeong:2017mzv}. The uncertainty arises in the extrapolation of the electromagnetic structure function $F_2(x,Q^2)$ which depends on the Bjorken-$x$ and momentum transfer squared $Q^2$ in the $\ell^\pm A\to \ell^\pm X$ scattering process. The Bezrukov and Bugaev \cite{Bezrukov:1981ci} (\texttt{bb}) formula for $\beta^\ell_{\rm nuc}$ does not account for $Q^2$ dependence. Parameterizations by Abramowicz et al. \cite{Abramowicz:1991xz,Abramowicz:1997ms} (\texttt{allm}), Block et al. \cite{Block:2014kza} (\texttt{bdhm}),  and Capella et al. \cite{Capella:1994cr}
(\texttt{ckmt}) have different extrapolations of $F_2(x,Q^2)$ to small-$x$. 

\begin{figure}
     \centering
     \includegraphics[width=0.48\textwidth]{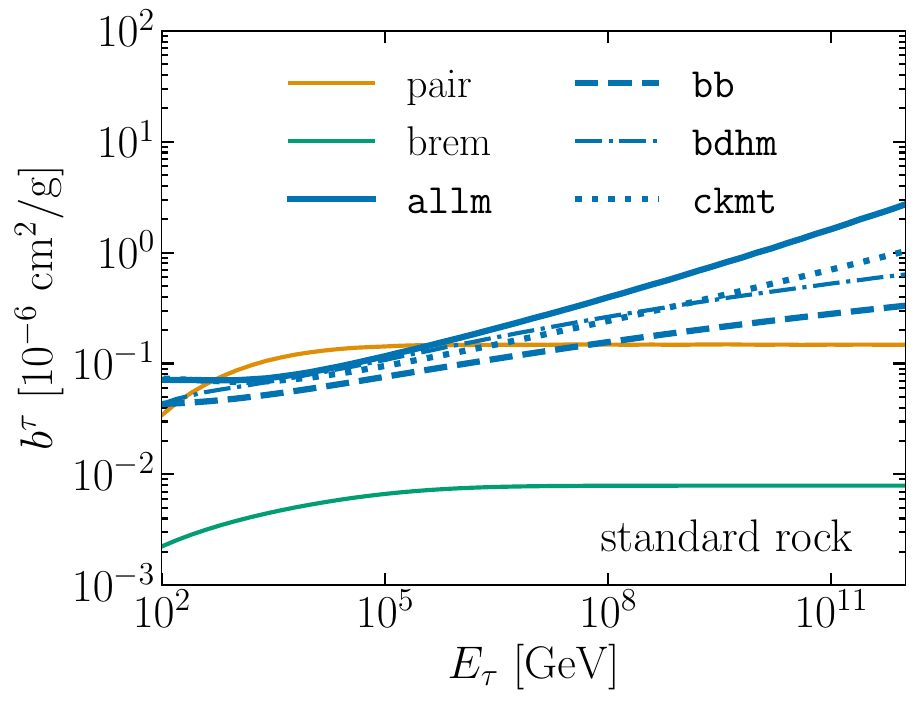}
     \includegraphics[width=0.48\textwidth]{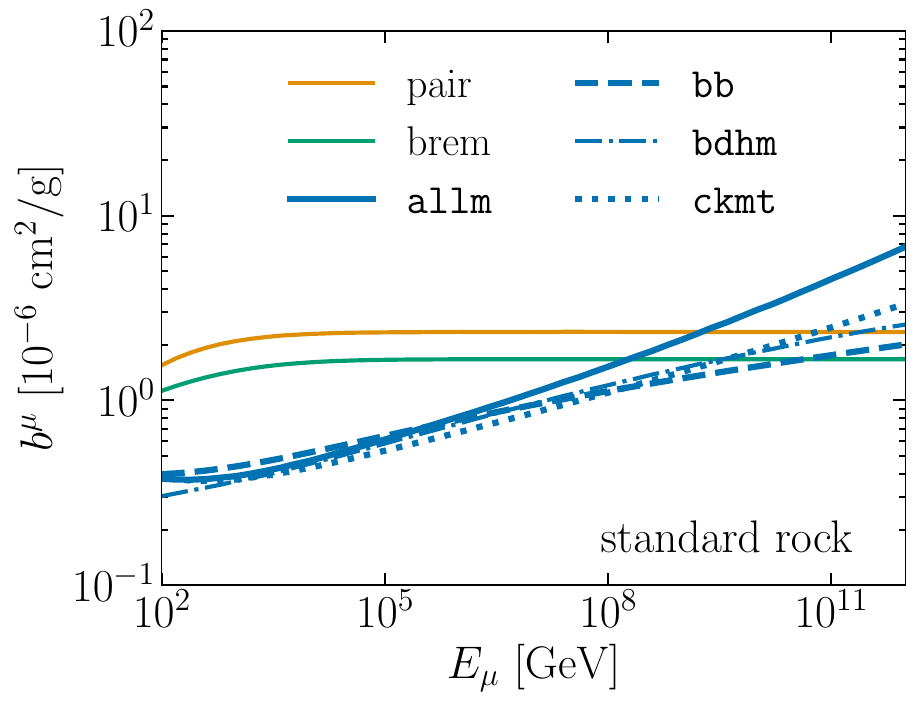}
     \caption{The electromagnetic energy loss parameters for \tauons\ (left) and muons (right) in rock. Photonuclear energy loss contributions from different high energy extrapolations are shown labeled by \texttt{allm} (Abramowicz et al. \cite{Abramowicz:1991xz,Abramowicz:1997ms}), \texttt{bb} (Bezrukov and Bugaev \cite{Bezrukov:1981ci}), \texttt{bdhm}  (Block et al. \cite{Block:2014kza}), and \texttt{ckmt} (Capella et al. \cite{Capella:1994cr}). Note the different $y$-axis scales in the two plots.}
     \label{fig:betas}
     \vspace{-10pt}
\end{figure}

The plots in \cref{fig:betas} show values of $b^\ell_i$ for \tauons\ and muons for standard rock for which $(Z, A) = (11, 22)$. Because $b^\ell_{\rm brem}$ scales approximately as $1/m_\ell^2$ (see, e.g., ref. \cite{Nachtmann:1990ta}), bremsstrahlung is significantly more important for muons than for \tauons. Pair production and photonuclear $b^\ell_i$ scale like $1/m_\ell$ (see, e.g., \cite{Tannenbaum:1990ae,Reno:2005si}), so electromagnetic energy loss for \tauons\ is lower than for muons. Note the different $y$-axis scales in \cref{fig:betas}.

The simulation of charged lepton energy loss depends on the electromagnetic interactions prior to the lepton decay or exit. The \nupyprop\ code has the option to treat electromagnetic energy loss via a stochastic or continuous energy loss algorithm. 
Multiple studies have been done to compare the effects of charged lepton propagation with continuous energy loss and that with stochastic energy loss \cite{Lipari:1991ut,Huang:2003,Dutta:2005yt,Bigas:2008ff,Dunsch:2018nsc}. Stochastic losses account for energy fluctuations in radiative processes \cite{Born:1939} and allow for catastrophic energy losses, whereas continuous energy losses are essentially treated as constants in small steps of propagation. The range of a charged lepton can be understood as the effective distance it propagates in a medium before it either decays or the if its energy drops below a specified threshold value. As the energy loss parameter ($b^\ell$) decreases as the mass of the charged lepton increases, muons undergo larger energy losses compared to $\tau$-leptons. The decay length of muons is far greater than that for $\tau$-leptons, so a muon will encounter far more energy loss interactions. Stochastic processes should play a more significant role in muons than $\tau$-leptons. 

Lipari and Stanev \cite{Lipari:1991ut} report a significant decrease in the range and survival probabilities of upward going muons when they are treated with stochastic energy losses compared to continuous energy losses. This  translates to muon exit probabilities that are lower when the more physical stochastic energy loss is implemented compared to when continuous energy loss is implemented for muons.  For \tauons, stochastic and continuous energy losses yield nearly the same results. We describe the two implementations here.

\begin{table}[]
\begin{tabular}{|c|c|c|c|}
    \hline
    & \bf{Energy Loss Process}    & $\mathbf{y_{min}}$ & $\mathbf{y_{max}}$ \\ \hline
& Bremmstrahlung  & 0       &  $10^{-3}$      \\ \hline
   \bf{Stochastic}  & Pair Production & $\frac{4 m_l}{E}$       & $10^{-3}$       \\ \hline
    & Photonuclear    &  $\frac{(m_p + m_\pi)^2 - m_p^2}{2 m_p E}$      & $10^{-3}$       \\ \hline
    \hline
 & Bremmstrahlung  & $10^{-7}$       &  1      \\ \hline
\bf{Continuous}  & Pair Production &  $\frac{4 m_l}{E}$      & $1 - \frac{3 m_l}{E} \sqrt{e} Z^{1/3}$       \\ \hline
 & Photonuclear    &  $\frac{(m_p + m_\pi)^2 - m_p^2}{2 m_p E}$      & $1 - \frac{m_l}{E}$       \\ \hline
\end{tabular}
\caption{\label{table:beta_y_vals} The minimum and maximum inelasticity $y$ for $\beta_{cut}$ for stochastic energy loss and $\beta$ for continuous energy loss. }
\end{table}

\subsection{The \texttt{propagate\_lep\_water} and \texttt{propagate\_lep\_rock} subroutines}

Depending on the location of the charged particle inside the Earth, we divide the propagation part of charged leptons in the code into two parts - propagation in water and propagation in all other materials, denoted ``rock." 
The \texttt{propagate\_lep\_water} subroutine applies to the trajectory of a charged lepton in the thin water layer that surrounds the Earth.  The main function of this subroutine is to return the final particle decay flag, the distance (in kmwe) the charged lepton travels through the water layer and the final energy of the charged lepton should it emerge. 
The \texttt{propagate\_lep\_rock} subroutine
is nearly identical to the \texttt{propagate\_lep\_water} subroutine. In this routine,  we include the calculation of the density of the material based on the Earth emergence angle and location along the trajectory where the interaction occurs, since the density changes based on where the charged lepton is inside the Earth. In \cref{app:z2a}, we describe our density dependent scaling of the values for $b_i^\ell$ for Earth densities between rock and iron, namely, for
$\rho_{rock} < \rho < \rho_{Fe}$. 
The one-dimensional approximation of the charged lepton trajectory is reliable at high energies. For final muon energies of order $10^3$ GeV, angular corrections may be as large as $0.5^\circ$, however, the median accummulated angular deviation is less than $0.1^\circ$, and decreases as the final muon energy increases \cite{Gutjahr:2022quk}. For \tauons , energies are larger than $\sim 10^5$ GeV where the one-dimensional approximation works well.

\subsubsection{Stochastic Losses}
\label{sec:stoch_losses}

To propagate a charged lepton using stochastic energy losses, we use the procedure described in refs.  \cite{Lipari:1991ut,Antonioli:1997qw}. The structure of our code follows that of ref. \cite{Antonioli:1997qw}. In this procedure, the energy loss parameter $b^\ell$ is split into two regimes, a soft term ($b^\ell_{cut}$), for which we treat the energy loss as a continuous process with inelasticity parameter $y_{min} < y \leq y_{cut}$, and a hard term which is responsible for stochastic loss:
\begin{eqnarray}
\label{eq:betacut}
    b_i^\ell(E) &=& \frac{N_A}{A} \Biggl[
    \int_{y_{min}}^{y_{cut}} y\ \frac{d\sigma_i^{\ell N}(y,E)}{dy}\ \mathrm{d}y+\int_{y_{cut}}^{y_{max}} y\ \frac{d\sigma_i^{\ell N}(y,E)}{dy}\ \mathrm{d}y\Biggr]\,\\ \nonumber
    &=& b_{i,cut}^\ell + 
    \frac{N_A}{A} \int_{y_{cut}}^{y_{max}} y\ \frac{d\sigma_i^{\ell N}(y,E)}{dy}\ \mathrm{d}y\,.
\end{eqnarray}
The \nupyprop\ code thus has lookup tables for $b^\ell_{cut}=\sum_i b^\ell_{i,cut}$ for $y<y_{cut}$ and separate tables for cross sections and energy distributions for $y_{cut} < y \leq y_{max}$. 

The separation between continuous energy loss for $y<y_{cut}$ (where the energy loss of the charged lepton is small) and stochastic energy loss otherwise helps to reduce the computation time of the code. The choice for $y_{cut}$ has been discussed in detail in many references \cite{Lipari:1991ut,Sokalski:2000nb,Dunsch:2018nsc}. We choose $y_{cut} = 10^{-3}$ and treat ionization loss continuously. 

With this implementation, the starting point for charged lepton propagation is to generate an interaction point, now based on the interaction length
\begin{equation}
\label{eq:int_depth_lep}
    X_{int} = \Bigl(\frac{N_A}{A}\,\sigma_{em}^{\ell A}(y>y_{cut}) +   \frac{1}{D_{dec}}\Bigr)^{-1} \
\end{equation}
where $D_{dec} = (\rho c \tau_\ell {E}_\ell/{m_\ell})$ is the time dilated decay length for the relativistic charged lepton in a material with density $\rho$. 

With the step size $X$ determined according to an exponential distribution determined by $\exp(-X/X_{int})$ for $X_{int}$ in \cref{eq:int_depth_lep}, continuous energy loss is applied using $b^\ell_{cut}$ and $a^\ell$. We omit the charged lepton and energy labels on $b^\ell_{cut}$ and $a^\ell$ in the equations below.
To first approximation, the continuous energy loss can be applied using the approximation
\begin{equation}
\label{eq:eavgapprox}
    \frac{dE}{dX}\simeq \Biggl\langle \frac{dE}{dX}\Biggr\rangle\,,
\end{equation}
with $b^\ell\to b^\ell_{cut}$.
Using \cref{eq:eavgapprox}, after a distance $X$, continuous energy loss 
for an initial charged lepton energy $E_i$ gives the energy $E_{cont}$ after continuous losses are included as
\begin{equation}
\label{eq:cont_loss}
    E_{cont} = E_i\ e^{-b^\ell X} - \frac{a^\ell}{b^\ell}(1 - e^{-b^\ell X})\,,
\end{equation}
with $a^\ell$ and $b^\ell=b^\ell_{cut}$ evaluated at $E_i$. Given that $b^\ell$ depends on energy, we make an adjustment that follows the procedure in ref. \cite{Antonioli:1997qw}, namely finding $E_{cont}'$ using $b^\ell_{cut}(E_{cont})$, then taking a logarithmic average 
\begin{equation}
    \label{eq:logavg}
    \log_{10}(E_{avg}/{\rm GeV}) = \frac{1}{2}\Bigl(\log_{10}(E_{cont}/{\rm GeV})+\log_{10}(E_{cont}'/{\rm GeV})\Bigr)\,.
\end{equation}
It is with $E_{avg}$ that we determine $a^\ell_{avg}$ and $b^\ell_{avg}$ to get
\begin{equation}
\label{eq:cont_loss}
    E_{int} = E_i\ e^{-b^\ell_{avg}X} - \frac{a^\ell_{avg}}{b^\ell_{avg}}(1 - e^{-b^\ell_{avg} X})\,.
\end{equation}
Using lookup tables for the bremsstrahlung, pair production and photonuclear cross sections and CDFs and using the time dilated lifetime for $E_{int}$, the interaction type or decay is determined, and for interactions, the energy of the lepton after the interaction is 
   \begin{equation}
        E_{lep} = E_{int} (1-y)
    \end{equation}
The loop continues until one of the following three cases is True:
1) the charged lepton energy falls below $E_{min}^{lep} = 1000$ GeV;
2)  the charged lepton decays; or
3)  the trajectory exceeds the maximum column depth of either the water layer or ``rock'' layer set by the PREM model. 

As discussed above, we use intermediate energy $E_{int}$ instead of the initial energy to determine the interaction type or decay. The energy dependence of the interaction part of $X_{int}$, $(N_A/A)\sigma_{em}^{\ell A}(y>y_{cut})$ in eq.~\eqref{eq:int_depth_lep}, is not very significant for determining the step size $X$. The effect is small because $\sigma_{em}^{\ell A}(y>y_{cut})$ (for $y_{cut}=10^{-3}$) doesn't change rapidly as a function of energy. For an electromagnetic interaction of {\tauon} of energy between 10$^7$ and 10$^9$ GeV, $\sigma_{em}^{\ell A}(y>y_{cut})$ increases by only 20\%. For the muons, the change is even smaller, $<$5\%. One aspect of our approach is the choice of $y_{cut}=10^{-3}$. As noted in ref.~\cite{Antonioli:1997qw}, the choice of $y_{cut}=10^{-2}$ for muon energy loss increases  $X_{int}$ by a factor of $\sim$ 10. 

\subsubsection{Continuous Losses}
\label{sec:cont_losses}

For continuous energy loss, we use the completely integrated form of the energy loss parameter ($b^\ell$) in \cref{eq:beta}. In this form of propagation, the column depth steps are calculated based on a fixed step size that we have set to 4500 cm. The smaller the step size, the longer is the computation time for the code. Our choice of the step size is justified by looking at the decay lengths of muons and $\tau$-leptons and accounting for a smoother transition from the water layer to the rock layer inside the Earth. We did not find any significant differences in the ranges, the $P_{exit}^{(\tau)}$ values or the outgoing energy distributions of the \tauons\ when we decreased the step size for $6 < \log_{10}(E_{lep}/\text{GeV}) < 12$. 
At each step of the propagation, the probability that the charged lepton decays is given by:
\begin{equation}
    P\mathrm{(decay)} = 1 - \exp\Biggl( 
    -\frac{\Delta L}{\gamma c \tau}
\Biggr) \,,
\end{equation}
where $\Delta L$ is the step size. 
We then use the lookup tables for the values of $a^\ell$ and $b^\ell$ that is fully integrated from the $y_{min}$ to $y_{max}$ values shown in table \ref{table:beta_y_vals} to find the charged lepton energy after the step, $E_f$, with continuous energy loss for each step $\Delta X = \rho \Delta L$ and energy $E_i$ at the beginning of the step, as:
\begin{equation}
    \label{eq:e_fin_lep}
        E_f = E_i\ e^{-b^\ell\Delta X} - \frac{a^\ell}{b^\ell}\Bigl(1 - e^{-b^\ell \Delta X}\Bigr)\,.
    \end{equation}
The value of $b^\ell$ here reflects the average energy loss the charged lepton encounters while traversing each step inside the propagation medium. The iterative loop stops execution if the updated $E_f$ value reaches $E_{min}^{lep} = 1000$ GeV or if the charged lepton decays.

\subsection{Muons and \tauons\ in rock}

\begin{figure}
     \centering
        \includegraphics[width=0.48\textwidth]{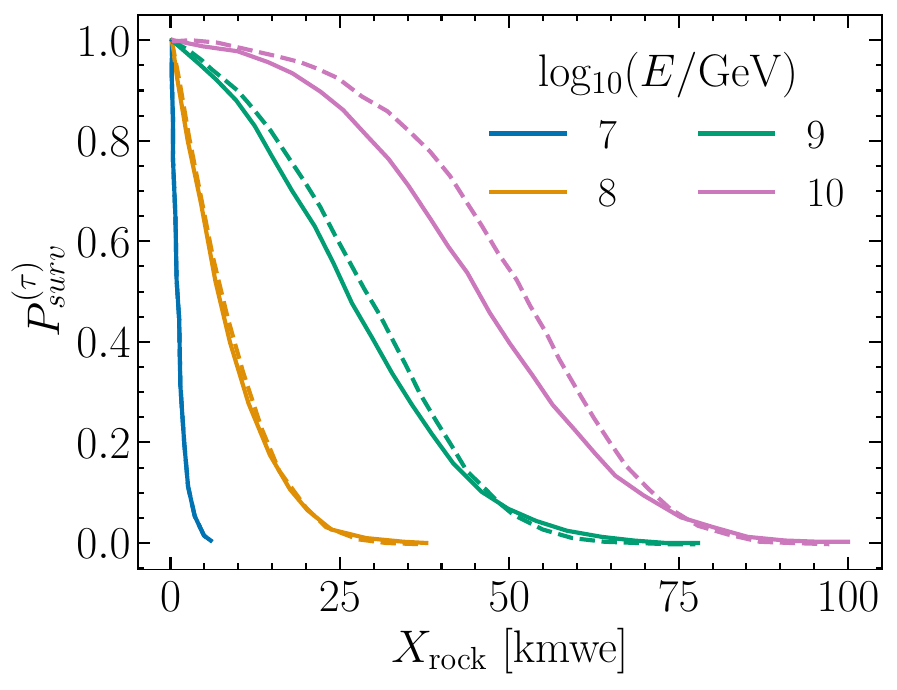}
     \includegraphics[width=0.48\textwidth]{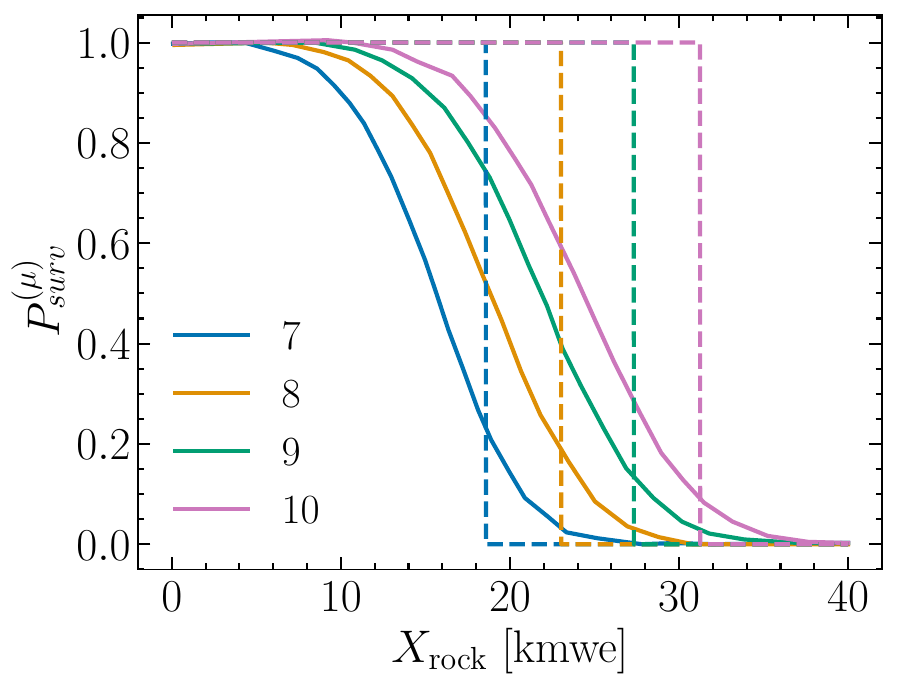}
     \caption{The \tauon\ and muon survival probabilities as a function of column depth $X_{\rm rock}$ in standard rock $(Z,A)=(11,22),\ \rho=2.65$ g/cm$^3$.
     The solid lines show stochastic energy loss results and dashed line are for continuous energy loss. The minimum charged lepton energy was taken to be $10^3$ GeV in the evaluation of the survival probabilities.}
     \label{fig:psurv_rock}
     \vspace{-10pt}
 \end{figure}

Comparisons of stochastic and continuous energy loss using \texttt{propagate\_lep\_rock} for \tauons\ and muons are shown in \cref{fig:psurv_rock}. The density of standard rock with $(Z,A)=(11,22)$ is $\rho=2.65$ g/cm$^3$. The survival probability as a function of column depth $d_{rock}$ is the fraction of charged leptons with fixed initial energies that survive to that distance, where both energy loss and the lifetime are included, until the the energy reduces to 
$10^3$ GeV. 

In the case of \tauons, at low energy, the column depth dependence is primarily dictated by the \tauon\ lifetime, while at higher energies, energy loss comes into play and there are some differences between evaluations with stochastic and continuous energy loss. The survival probability is not sensitive to the low minimum energy $10^3$ GeV because the decay length for $E_\tau=10^7$ GeV is approximately 450 m $\simeq 1.2$ km.w.e. column depth. The decay of \tauons\ means that its energy will not reduce to such a low energy as  $10^3$ GeV. 

On the other hand, for muons where $\gamma c \tau$ for $E_\mu=10^3$ GeV is $\sim 6000$ km, the minimum energy determines the survival probability. With continuous energy loss, using \cref{eq:eavgapprox} and \cref{eq:e_fin_lep} for each step $\Delta X$, the initial and final muon energies determine the distance travelled, hence the sharp cutoff in the survival probabilities. With stochastic energy loss, fluctuations in the energy loss produce a distribution for the survival probabilities as seen in the right plot of \cref{fig:psurv_rock}. 

\subsection{Regeneration in \tauon\ decays}
\label{sec:regeneration}

With a lifetime much shorter than for muons, \tauon\ decay is an important feature of simulations of \tauon\ propagation in the Earth. When they decay, they produce tau neutrinos, so each \tauon\ decay regenerates a tau neutrino some distance from where the tau neutrino was absorbed. This can have a profound effect on the number of \tauons\ exiting the Earth as it has the ability to create a chain of tau neutrino and \tauons\ during propagation
\cite{Halzen:1998be,Iyer:1999wu,Dutta:2000jv,Becattini:2000fj,Bugaev:2003sw,Bigas:2008sw}.
Thus, the propagation with regeneration in \nupyprop\ follows the same logic as for incident neutrinos, using \texttt{propagation\_nu} and the charged lepton propagation code, now with a new incident neutrino energy started at the \tauon\ decay point on the trajectory in the Earth, with a new, lower neutrino energy that accounts for the fact that the decaying tau has lost energy prior to decay after having been produced in an earlier $\nu_\tau$ CC interaction, and that the energy of the neutrino from the decay comes from the decay distribution of the \tauon . The only new feature in regeneration is the determination of the decay $\nu_\tau$ energy.

A \tauon\ always produces a $\nu_\tau$. The other constituents of the decay product can be either leptons or hadrons. The different decay channels and their branching ratios for $\tau$-lepton decay are listed in table \ref{table:taudecays} in \cref{subsec:taudecay}. The CDF for the energy of the $\nu_\tau$ from the \tauon\ decay is used to determine the $\nu_\tau$ energy. We approximate the full decay CDF by the \tauon\ leptonic decay distribution for left-handed taus. Note that left-handed \tauon\ decays of $\tau^-$ yield the same $\nu_\tau$ energy distribution as the
$\bar\nu_\tau$ distribution from right-handed $\tau^+$ decays.

We justify these approximations in \cref{subsec:taudecay}. To a good approximation, the energy distribution of the tau neutrino from \tauon\ leptonic decays represents the energy distribution of all decay channels combined when finite width effects are included for semileptonic decays involving the $\rho$, $a_1$ and $4\pi$ in the final states. In \cref{subsec:taudecay}, we also discuss the role of the tau polarization in the energy distribution of the tau neutrinos from \tauon\ decays. We show in \cref{subsec:polarization} that the exit probabilities and energy distributions of the \tauons\ are not significantly affected by accounting for \tauon\ depolarization effects that occur in \tauon\ electromagnetic scattering, a feature discussed in more detail in ref. \cite{Arguelles:2022bma}.

\begin{figure}[H]
     \centering
     \includegraphics[width=0.7\textwidth]{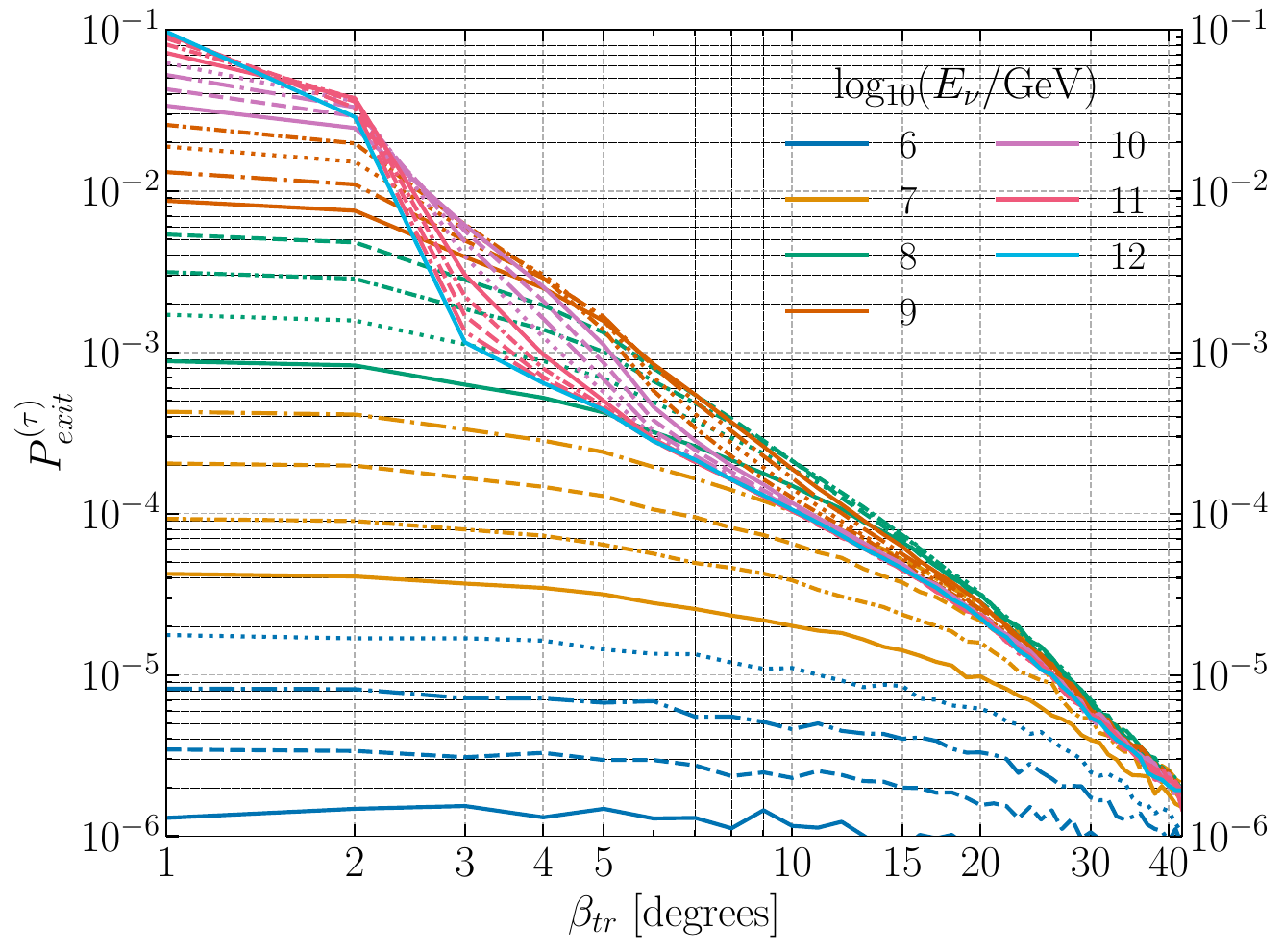}
     \caption{The tau exit probability versus Earth emergence angle ($\beta_{tr}$) from \nupyprop, for energies in steps of 0.1 in $\log_{10}(E_\nu/{\rm GeV})$, using default settings that include stochastic electromagnetic energy loss using the \texttt{allm} photonuclear energy loss inputs.}
     \label{fig:p-rainbow}
 \end{figure}

\section{Selected Results}
\label{sec:results}
\subsection{Exit probabilities and energy distributions from \nupyprop}

We show some representative results from \nupyprop. Unless specifically noted, we use our default inputs: 4 km of water in the outer layer of the Earth, the \texttt{ct18nlo} neutrino cross sections and the \texttt{allm} setting for charged lepton photonuclear interactions.
\Cref{fig:p-rainbow} summarizes our results for the tau exit probabilities as a function of Earth emergence angle $\beta_{tr}$ for energies in steps of 0.1 in $\log_{10}(E_\nu/{\rm GeV})$. The feature at high energies between $2^\circ-3^\circ$ comes from the transition from a column depth of all water to layers of rock and water, and at higher angles, more dense matter. This transition is less apparent in the other figures shown below as they come from evaluations using unit degree angular steps for the Earth emergence angle.

 \begin{figure}[H]
     \centering
     \includegraphics[width=0.7\textwidth]{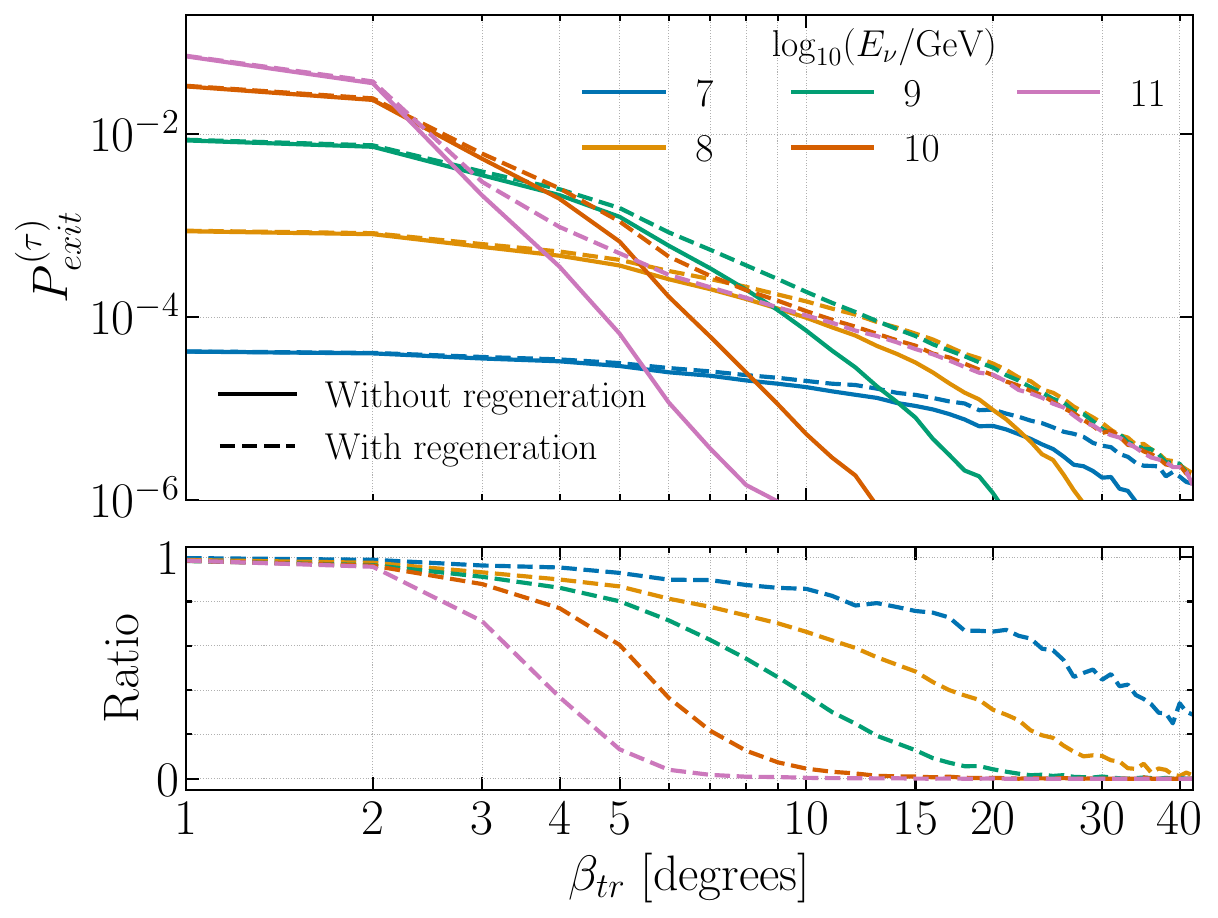}
     \caption{The tau exit probability versus Earth emergence angle ($\beta_{tr}$) with and without regeneration for incident $\nu_\tau$ energies of $E_\nu=10^7$, $10^8, \ \ldots, 10^{11}$ GeV.}
     \label{fig:p-regen}
 \end{figure}

\Cref{fig:p-regen} illustrates the impact of tau neutrino regeneration in \tauon\ propagation through the Earth. The lower panel of the figure shows the ratio of $P_{exit}^{(\tau)}$ without regeneration included to $P_{exit}^{(\tau)}$ with regeneration. At low angles, the ratio is one, while at higher angles, the ratio decreases. The decrease changes slowly as a function of $\beta_{tr}$ for incident neutrino energy $E_\nu=10^7$ because the \tauon\ lifetime is short. The \tauons\ that emerge for $E_\nu=10^7$ almost all come from the first $\nu_\tau$ CC interaction, an interaction that is near enough to the surface of the Earth so the \tauon\ can exit before it decays. On the other hand, for $E_\nu=10^{11}$ GeV and $\beta_{tr}=5^\circ$, the \tauon\ exit probability including regeneration is almost a factor of 10 larger than the case without regeneration. As angles increases, the number of regeneration steps increases, as has been emphasized in ref. \cite{Alvarez-Muniz:2018owm}.

\begin{figure}[H]
     \centering
     \includegraphics[width=0.48\textwidth]{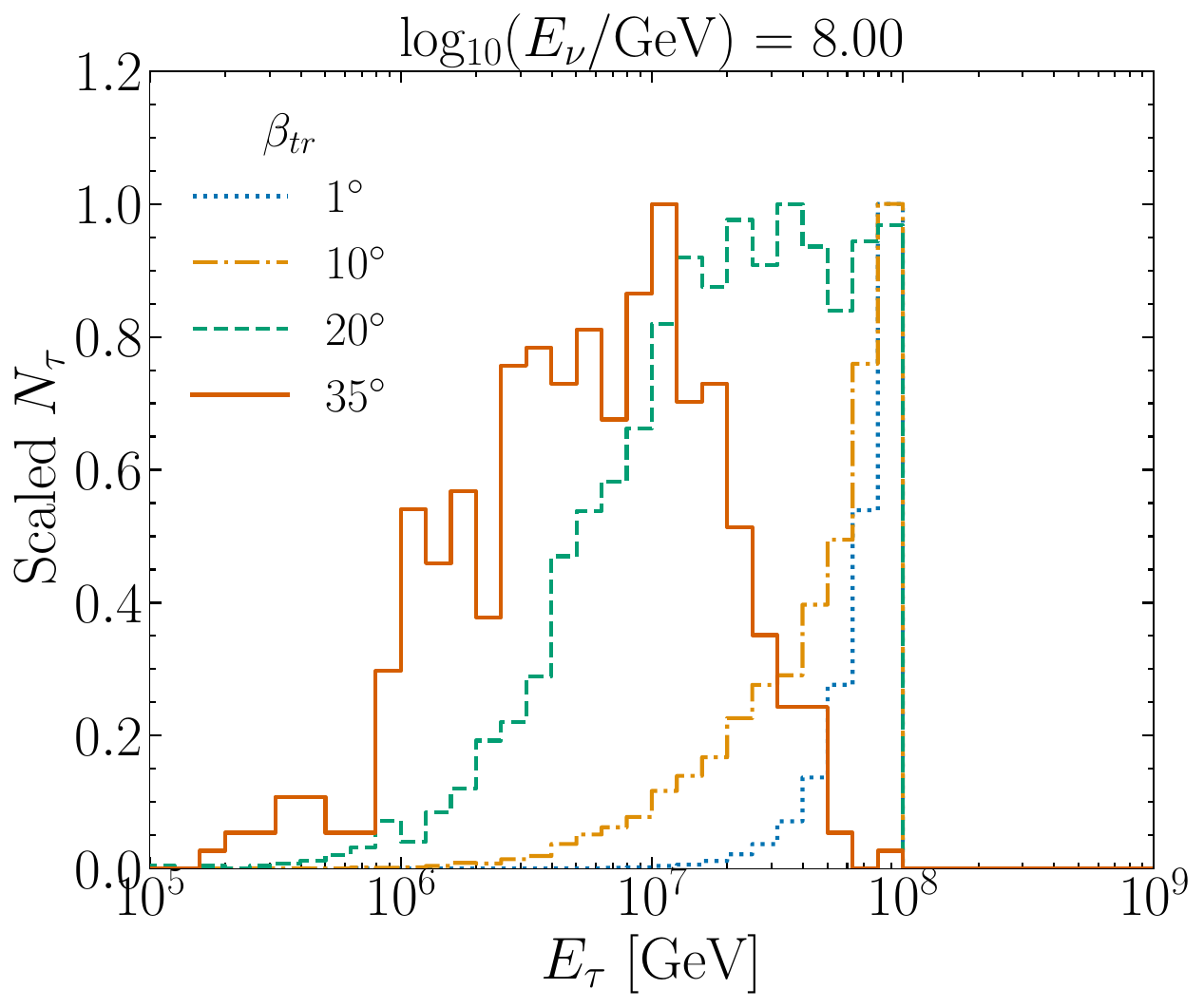}
\includegraphics[width=0.48\textwidth]{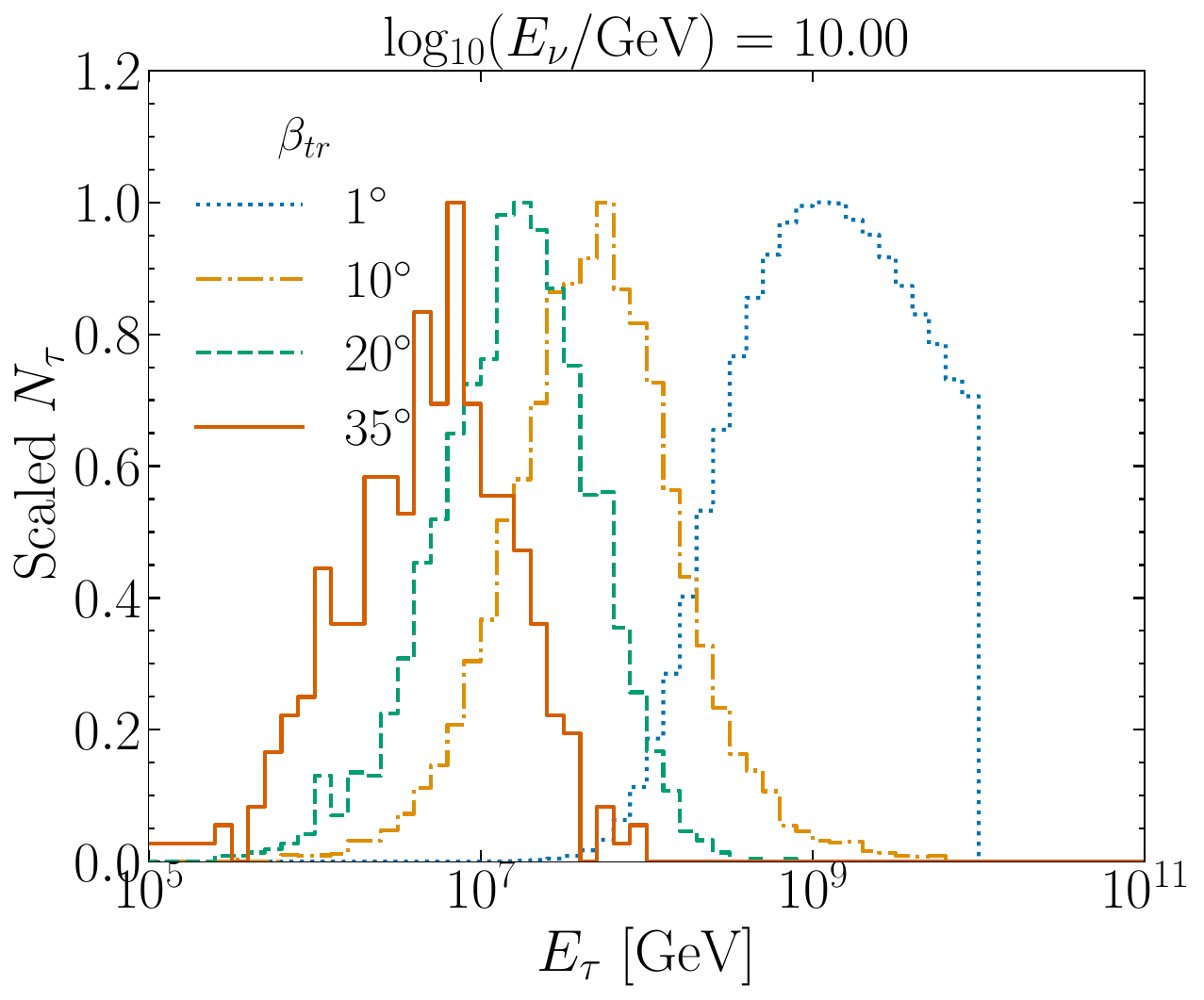}     
     \caption{The tau energy distributions for $\beta_{tr}=1^\circ,\ 10^\circ,\ 20^\circ,\ 35^\circ$ for incident neutrino energies $E_\nu=10^8$ GeV (left) and $E_\nu=10^{10}$ GeV (right).}
     \label{fig:energy-dist-8-10}
 \end{figure}

The output CDFs for the emerging \tauon\ energies come from the \tauon\ energy distribution, shown in \cref{fig:energy-dist-8-10} for $E_\nu=10^8$ GeV (left) and 
$E_\nu=10^{10}$ GeV (right). For $E_\nu=10^8$ GeV, the increase in Earth emergence angle shifts the peak of the distribution of outgoing energies to lower energies. Only for the largest angles is regeneration important. The right panel of \cref{fig:energy-dist-8-10} shows a much more abrupt shift in the position of the peak in the step from $\beta_{tr}=1^\circ$ to $\beta_{tr}=10^\circ$. The high energy peak in exiting \tauon\ energy distribution for $\beta_{tr}=1^\circ$ occurs because most of the exiting \tauons\ come from the first $\nu_\tau$ CC interaction. Energy loss of the \tauons\ over the long decay lengths of such high energy \tauons\ allow for energy losses  that shift the peak of the exiting \tauon\ energy distribution to close to $0.1\times E_\nu$ for $E_\nu=10^{10}$ GeV and $\beta_{tr}=1^\circ$. For larger Earth emergence angles, regeneration effects are also important, so the peaks of the exiting \tauon\ energy distributions are more significantly shifted to lower energies.
 
 \begin{figure}[H]
     \centering
     \includegraphics[width=0.8\textwidth]{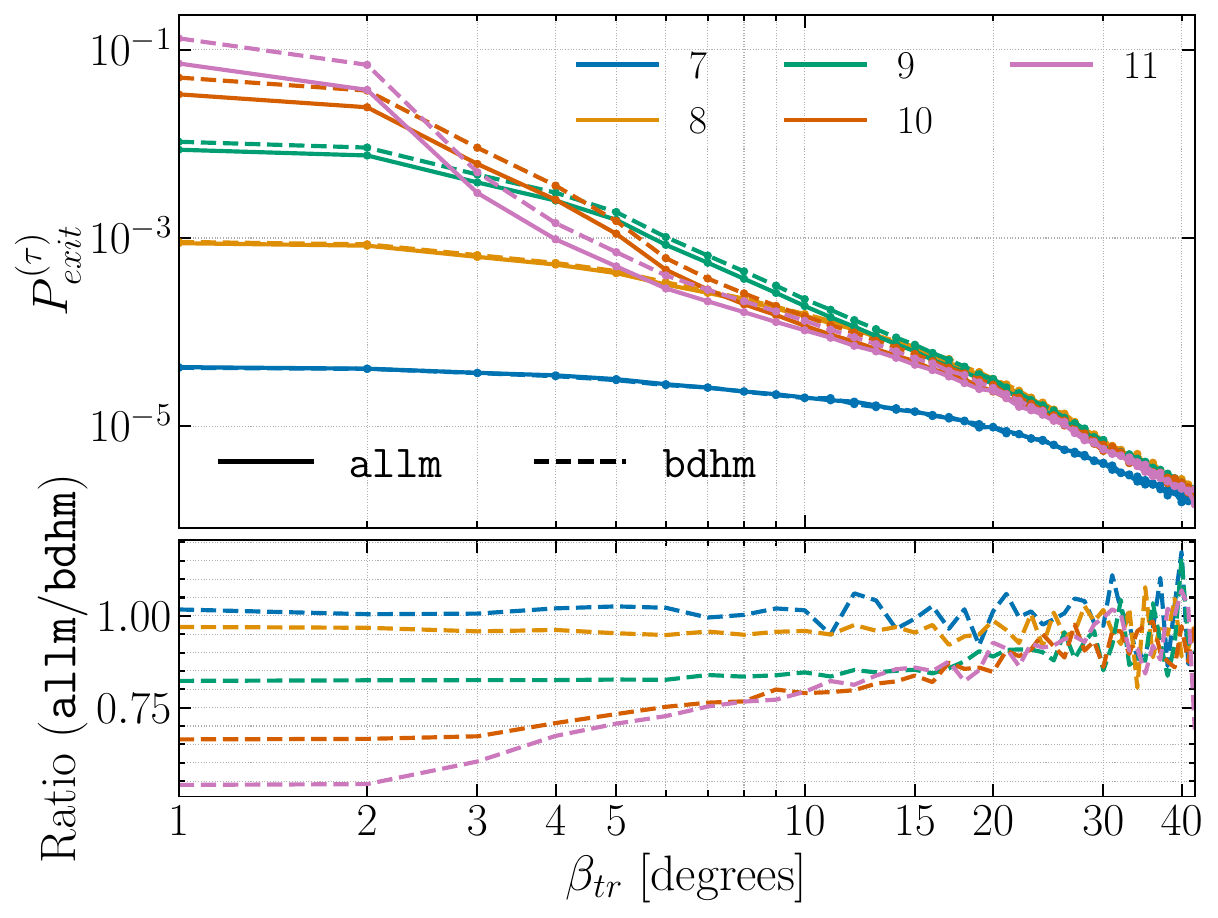}
     \caption{The tau exit probability versus Earth emergence angle ($\beta_{tr}$) using \texttt{allm} and \texttt{bdhm} parameterizations of the photonuclear energy loss and the ratio of the exit probabilities using \texttt{allm} to \texttt{bdhm}.}
     \label{fig:p-allm-bdhm}
 \end{figure}

\begin{figure}[H]
     \centering
     \includegraphics[width=0.9\textwidth]{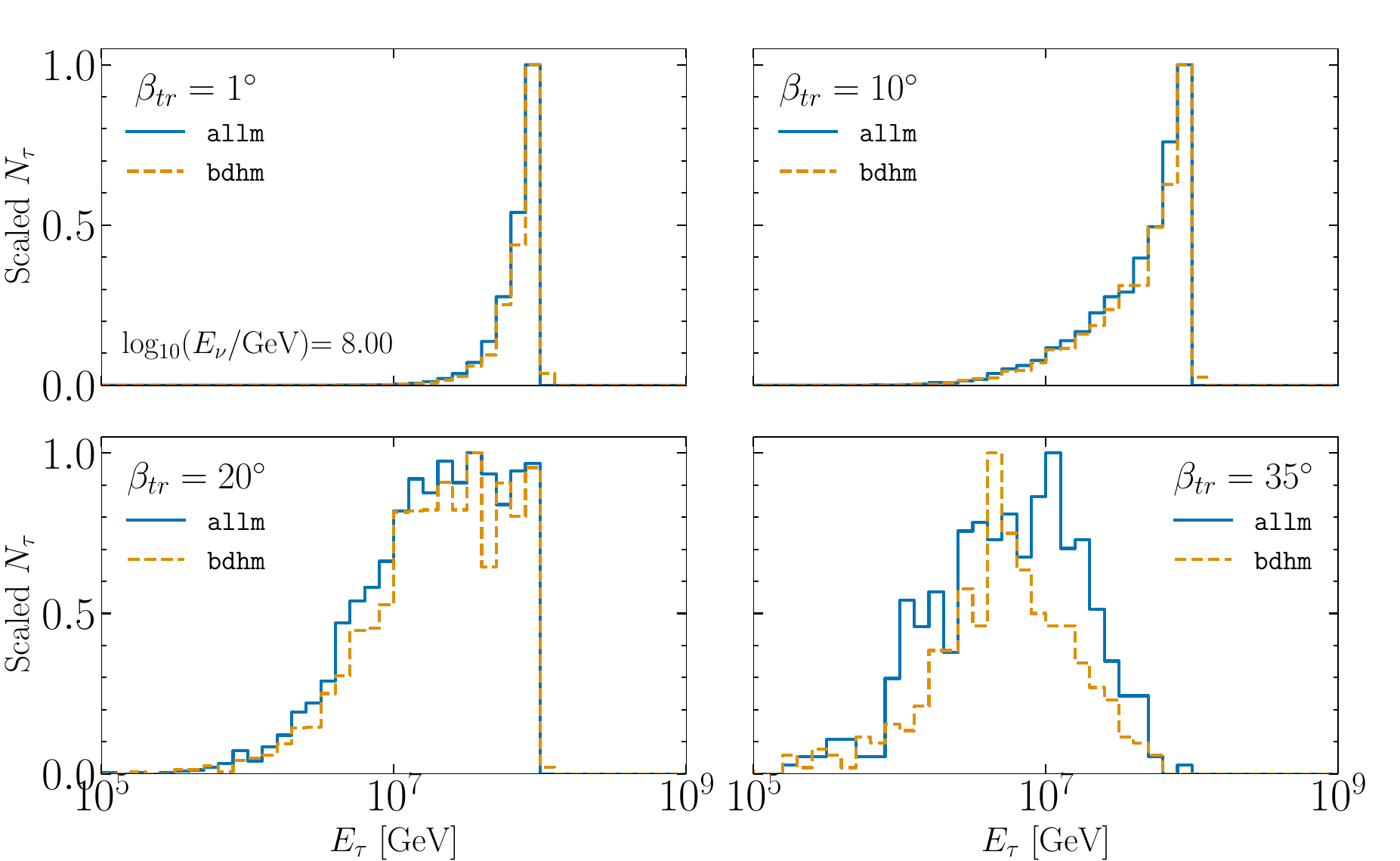}
     \caption{The energy distributions of \tauons\ that exit the Earth for $\beta_{tr}=1^\circ, 10^\circ, 20^\circ, 35^\circ$ given incident tau neutrinos, $E_\nu=10^8$ GeV, evaluated using the \texttt{allm} (default) and \texttt{bdhm} parameterizations of $F_2$ in the photonuclear energy loss evaluation.}
     \label{fig:edist-8-pnmodels}
 \end{figure}

\begin{figure}[H]
     \centering
     \includegraphics[width=0.9\textwidth]{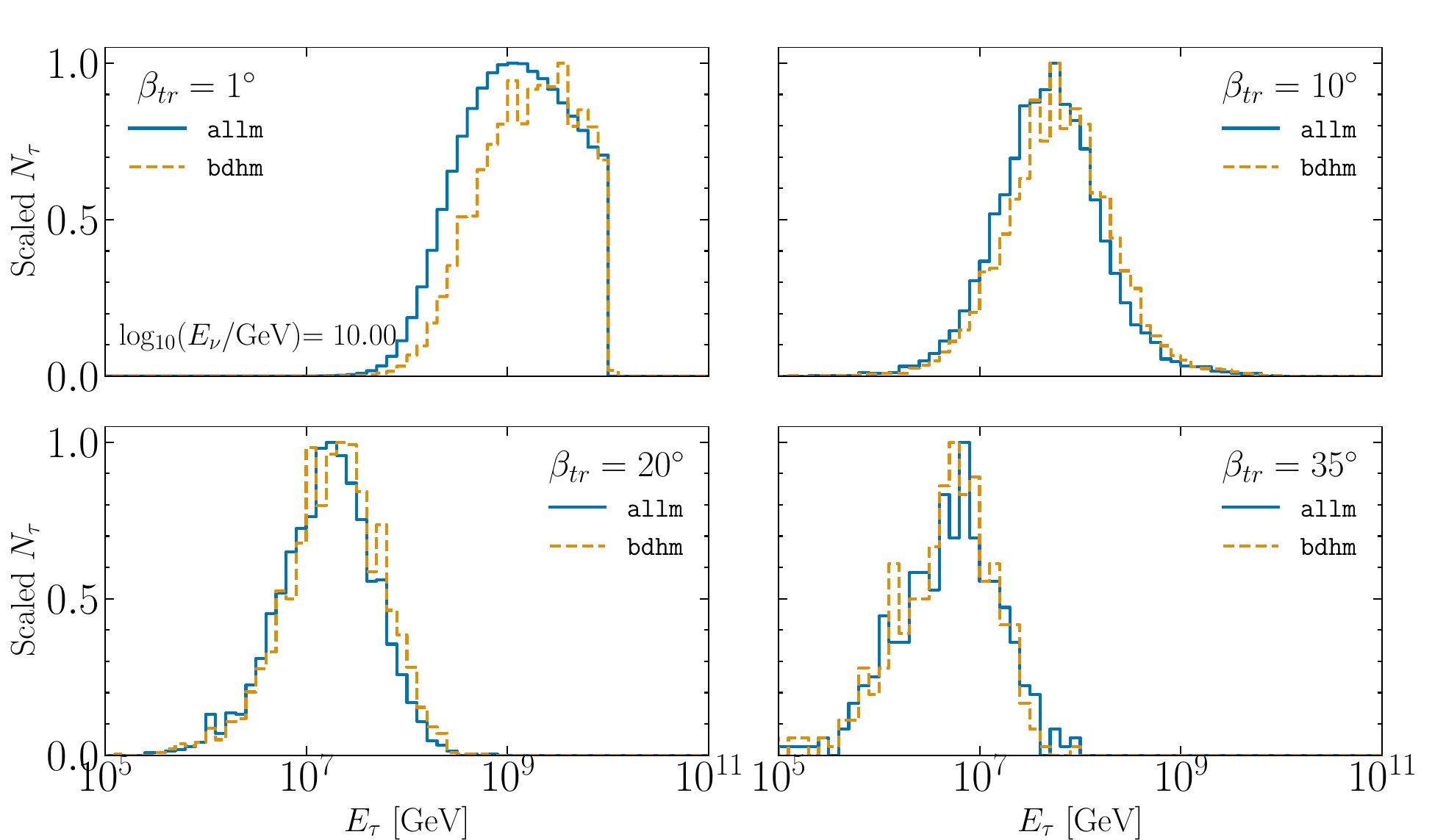}
     \caption{The energy distributions of \tauons\ that exit the Earth for $\beta_{tr}=1^\circ, 10^\circ, 20^\circ, 35^\circ$ given incident tau neutrinos, $E_\nu=10^{10}$ GeV, evaluated using the \texttt{allm} (default) and \texttt{bdhm} parameterizations of $F_2$ in the photonuclear energy loss evaluation.}
     \label{fig:edist-10-pnmodels}
 \end{figure}

In \cref{fig:p-allm-bdhm}, we compare the \tauon\ exit probabilities evaluated using the \texttt{allm} (default) and \texttt{bdhm} parameterizations of $F_2$ in the photonuclear energy loss evaluation. For $E_\nu=10^7$ GeV, \tauon\ decays determine the exit probability rather than energy losses so the ratio of $P_{exit}^{(\tau)}$ using \texttt{allm} to that using \texttt{bdhm} is nearly unity. For higher energies, the larger $\beta^\tau_{\rm nuc}$ for \texttt{allm} compared to 
$\beta^\tau_{\rm nuc}$ for \texttt{bdhm} means that fewer \tauons\ exit in the \texttt{allm} evaluation than for \texttt{bdhm}. For example, for $E_\nu=10^{10}$ GeV, the ratio is $\sim 0.7$. While the exit probabilities show some differences, the energy distributions of the exiting \tauons\ show less of an effect, as illustrated in~\cref{fig:edist-8-pnmodels} and~\cref{fig:edist-10-pnmodels}.

\subsection{Stochastic versus continuous energy losses}

\begin{figure}[H]
     \centering
     \includegraphics[width=0.7\textwidth]{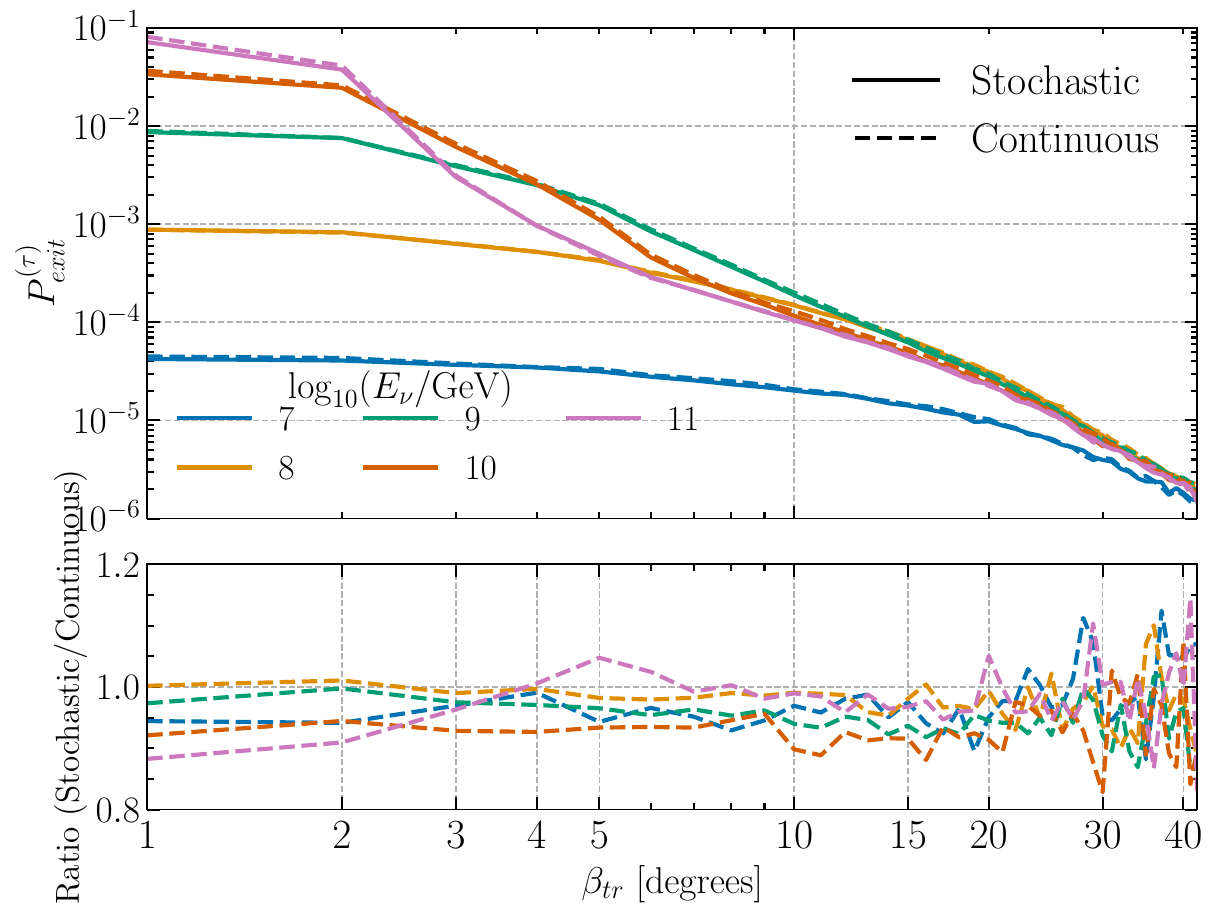}
     \caption{The tau exit probability versus Earth emergence angle ($\beta_{tr}$) from \nupyprop\ using stochastic (solid) and continuous (dashed) electromagnetic energy loss (upper) and the ratio of $P^{(\tau)}_{exit}$ evaluated with stochastic to continuous energy loss (lower).}
     \label{fig:p-stoch-cont}
     \vspace{-10pt}
 \end{figure}

A comparison between continuous and stochastic energy losses for the tau exit probability are shown in 
\cref{fig:p-stoch-cont}. For \tauons, the results differ by $\sim 10\%$ or less for most energies and Earth emergence angles.  While stochastic energy loss is more physical and allows one to include polarization effects, \cref{fig:p-stoch-cont} shows that an implementation of either stochastic or continuous \tauon\ energy loss yields comparable results. The out-going \tauon\ energy distributions are also comparable.

 \begin{figure}[]
     \centering
     \includegraphics[width=0.7\textwidth]{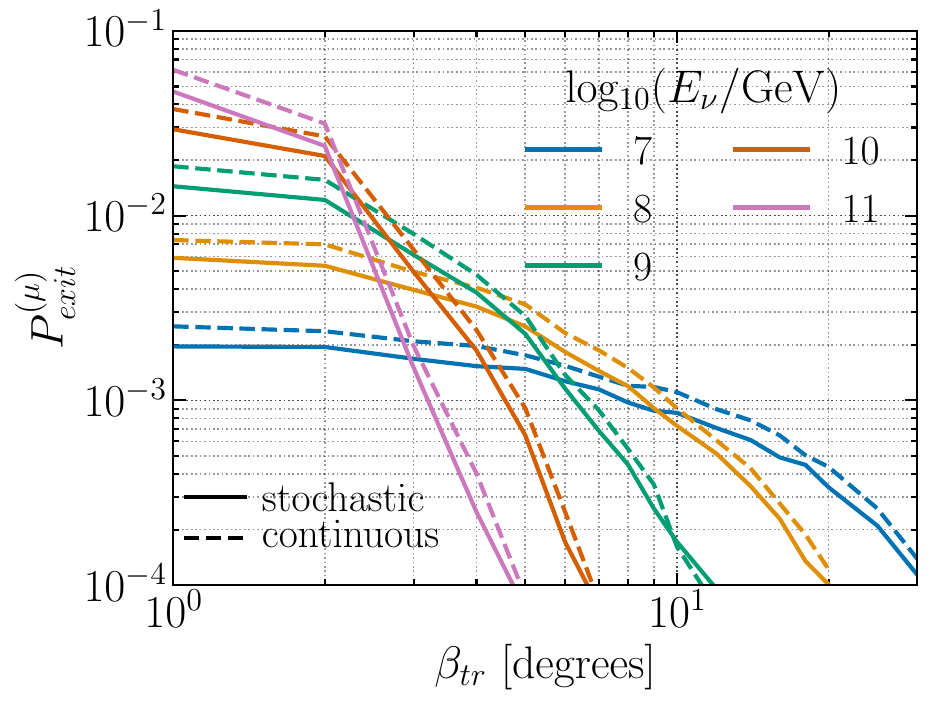}
     \caption{The muon exit probability versus Earth emergence angle ($\beta_{tr}$) from \nupyprop\ using stochastic and continuous electromagnetic energy loss. The ratio of stochastic to continuous is  $\sim 0.8$ for the angles and energies shown here.
     The minimum muon energy is set to $10^3$ GeV.}
     \label{fig:p-muon-stoch-cont}
     \vspace{-10pt}
 \end{figure}
 
For muon production and propagation, the implementation of stochastic energy loss has a much larger impact, as shown in \cref{fig:p-muon-stoch-cont}. 
Stochastic energy loss results for $P^{(\mu)}_{exit}$ are $\sim 20\%$ lower than when evaluated using continuous energy loss. This is not surprising given the discrepancies in the muon survival probabilities as a function of distance shown in the right panel of \cref{fig:psurv_rock}, and the fact that there is no energy smearing from  $\nu_\mu\to\mu\to\nu_\mu$ regeneration since few muons decay in the energy ranges considered here.

\subsection{Comparisons with other propagation codes}

A number of other codes propagate tau neutrinos and muon neutrinos through the Earth. They include \texttt{NuPropEarth} \cite{Garcia:2020jwr}, \texttt{TauRunner} \cite{Safa:2021ghs}, \texttt{NuTauSim} \cite{Alvarez-Muniz:2018owm} and \texttt{Danton} \cite{Niess:2018opy}. Except for \texttt{NuTauSim}, the codes use stochastic energy losses. The  \texttt{NuPropEarth} codes includes sub-leading contributions to the neutrino cross sections at high energies. 
Some codes track other neutrino flavors as well as all charged leptons in addition to the charged lepton associated with the incoming neutrino flavor. A more detailed comparison of these codes appears in ref. \cite{Abraham:2022jse}.

\begin{figure}[H]
     \centering
     \includegraphics[width=0.7\textwidth]{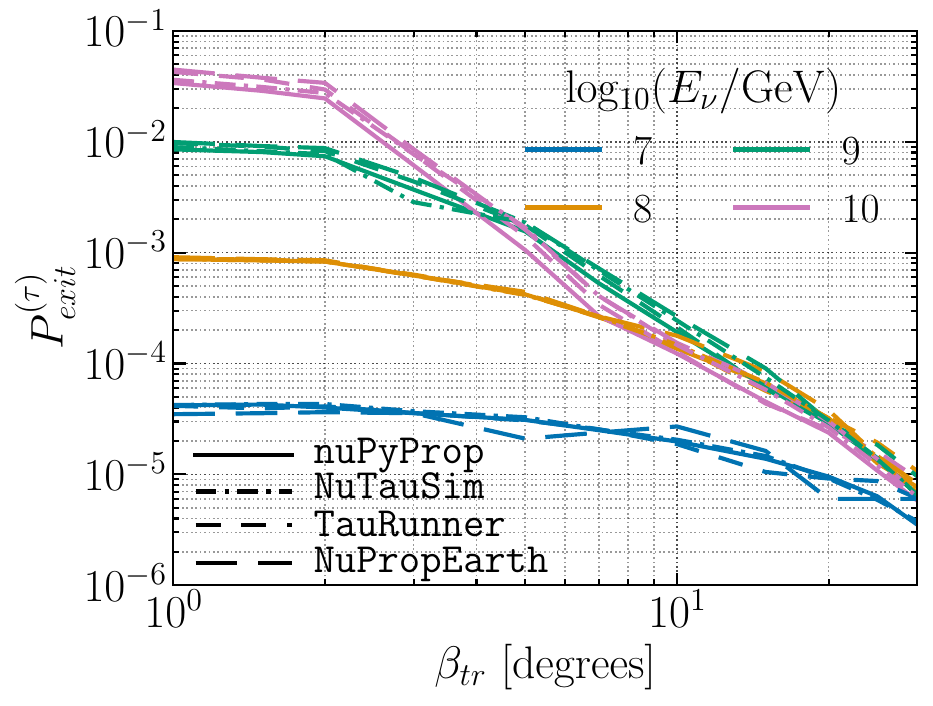}
     \caption{A comparison of the tau exit probabilities as a function of Earth emergence angle ($\beta_{tr}$) for  \nupyprop, \texttt{NuTauSim}, \texttt{TauRunner} and \texttt{NuPropEarth} propagation codes, given an Earth water layer of 4 km and with the \texttt{allm} photonuclear energy loss input. Figure as in ref. \cite{Abraham:2022jse}.}
     \label{fig:pexit-compare}
\end{figure}

Given different implementations of interactions, energy loss and tau decays, it is interesting to compare results for the \tauon\ exit probability as a function of energy and Earth emergence angle. \Cref{fig:pexit-compare} shows the \tauon\ exit probability for \nupyprop, \texttt{NuPropEarth}, \texttt{TauRunner} and \texttt{NuTauSim}, all run with a 4 km water layer and using the \texttt{allm} parameterization of the electromagnetic structure function input to the photonuclear cross section. There is reasonably good agreement between results across energies and angles.

We tested the speed of {\nupyprop} code for different neutrino energies and Earth emergence angles. For reference, a table is provided in the Readme file on the GitHub \cite{nupyprop_github} of {\nupyprop}. For example, for $E_\nu=10^7$ GeV and $\beta_{tr}=1^\circ-35^\circ$ it takes about one-hour to run the simulation for a computer system with 6 cores and 12 threads, and 26 minutes for a system with 8 cores and 16 threads.
 
\subsection{Modeling Uncertainties}
 
For the highest energy shown in fig. \ref{fig:pexit-compare}, differences in the high energy cross section explain some of the discrepancy between \nupyprop\ and {\texttt NuPropEarth}. The {\texttt NuPropEarth} code \cite{Garcia:2020jwr} includes subleading corrections to the neutrino cross section, for example, real $W$ and lepton trident production \cite{Zhou:2019frk}. Uncertainties in the high energy neutrino cross section that come from extrapolations of the parton distribution functions to kinematic ranges outside of current measurements begin at $E_\nu=10^8-10^9$ GeV and may increase to a cross section uncertainty as large as a factor of $2$ at $E_\nu=10^{12}$ GeV \cite{Connolly:2011vc,Garcia:2020jwr,Cooper-Sarkar:2011jtt}.
 
\begin{figure}[]
    \begin{center}
    \centering
    \includegraphics[width=0.7\textwidth]{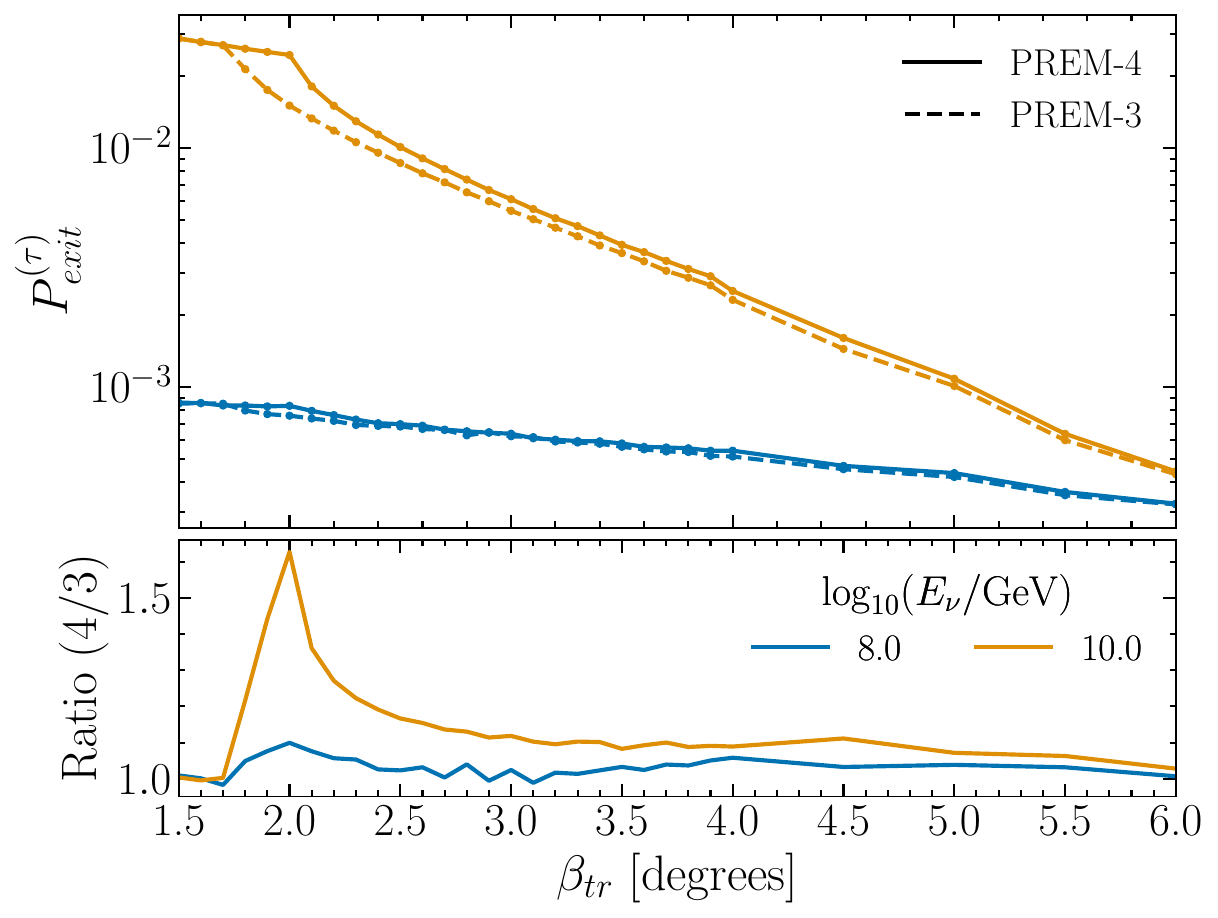}
    \includegraphics[width=0.7\textwidth]{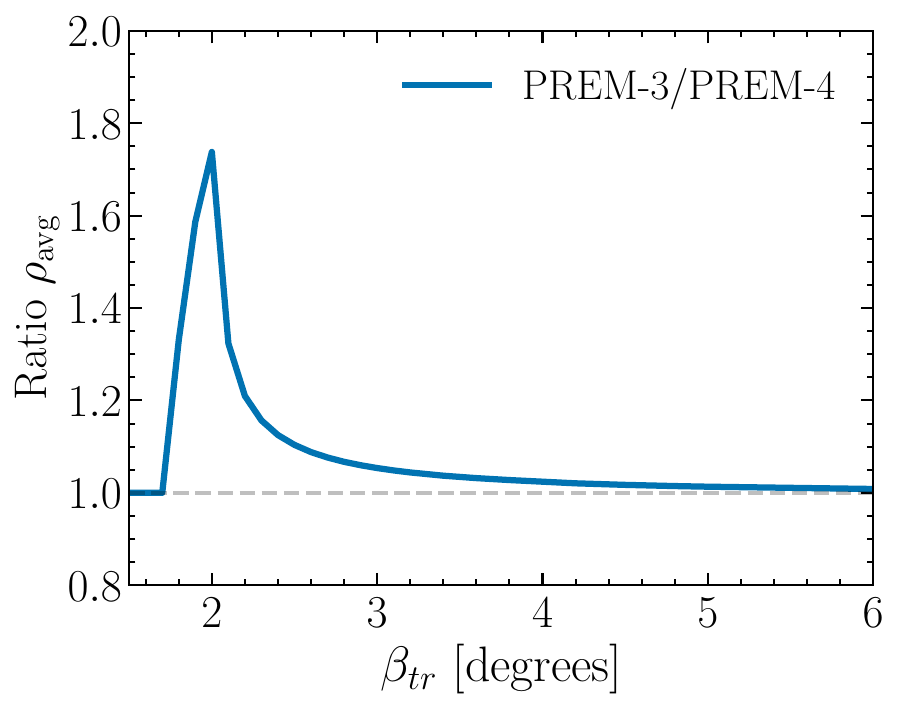}   
    \caption{The tau exit probabilities and ratio of the exit probabilities for 4 km and 3 km water depths in the PREM Earth model, and their ratios, as a function of Earth emergence angle for $E_\nu=10^8$ GeV and $10^{10}$ GeV (top). Also shown (bottom) is the ratio of the PREM (3 km water) and PREM (4 km water) average densities a function of Earth emergence angle for $\beta_{\rm tr}=1.5^\circ-6^\circ$.}
    \label{fig:pexit-34}
    \end{center}
\end{figure}

Modeling of the Earth's density distribution impacts results as shown in fig. \ref{fig:pexit-34}, where we show the exit probability for our standard water depth of 4 km and the water depth of the PREM model \cite{DZIEWONSKI1981297}. Below the water layer, both evaluations use the PREM density as a function of Earth radius. The average depth of the ocean is $\sim 3.7$ km. 
For $E_\nu=10^8$ GeV, the tau exit probability shows a variation of less than $10\%$ over the transition region where the densities differ. 
For a water depth of 3 km, trajectories with $\beta_{tr}>1.76^\circ$ traverse some rock, while for a water depth of 4 km, the critical angle beyond which the trajectory includes rock is $\beta_{tr}=2.03^\circ$. The shape of the ratio of the tau exit probabilities for 4 km and 3 km of water depth has the same shape and roughly the same magnitude as the inverse ratio of average densities along the same trajectories shown in the lower panel of fig. \ref{fig:pexit-34}. The extra rock for the 3 km case causes more energy loss for \tauons\  produced in rock, thus lowering the exit probability relative to the 4 km water case.

\begin{figure}[htbp]
\begin{center}
   \centering
\includegraphics[width=0.7\textwidth]{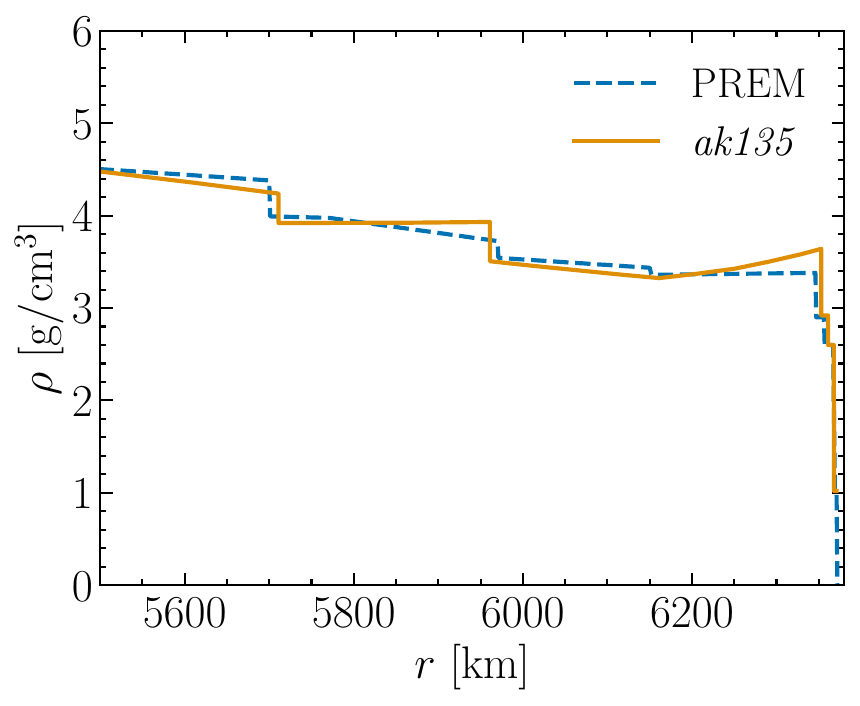}   
\includegraphics[width=0.7\textwidth]{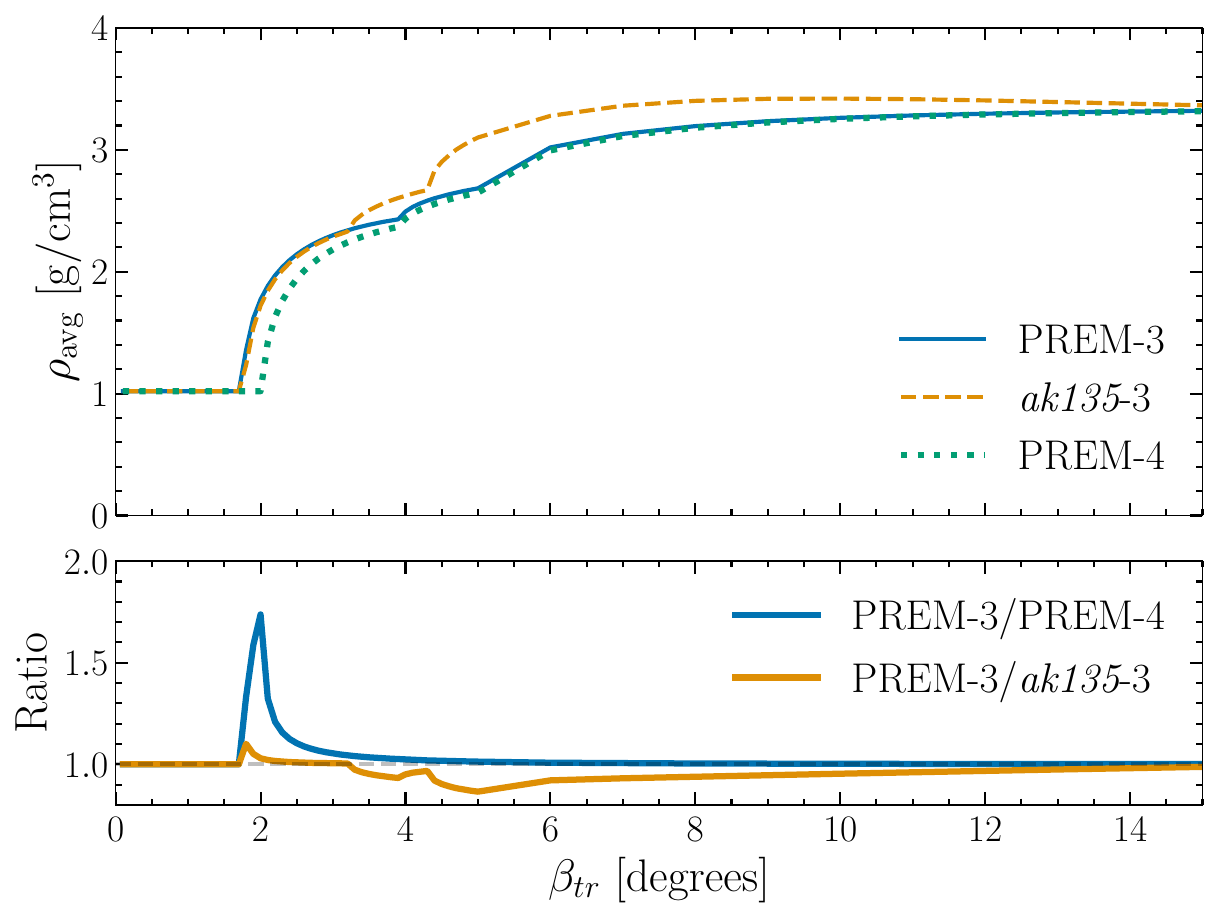}
\caption{Upper: The PREM and {\it ak135} Earth density model as a function of Earth radius. Both models have the water depth of 3 km.
Lower:The average density along a chord in the Earth as a function of the trajectory's Earth emergence angle $\beta_{tr}$ for the PREM-3 (same as PREM) and {\it ak135}-3 (same as {\it ak135}) Earth density models, and for the PREM density model modified to have a water depth of 4 km (PREM-4). Also shown  is the ratio of the PREM-3 and PREM-4 average densities and the ratio of the PREM to {\it ak135} (both with 3 km water) average densities along the chord.}
\label{fig:avgdensity}
\end{center}
\end{figure}
 
The PREM parameterization of the average Earth density as a function of radius is updated in the {\it ak135} Earth density model \cite{Kennett:1995ak}. Also with a water depth of 3 km, the {\it ak135} density model comes from improved analyses of seismic wave data. The upper panel of fig. \ref{fig:avgdensity} shows the density profile as a function of radial distance. The lower panel shows the average density along the chord labeled by $\beta_{tr}$ for the PREM model with water depths of 3 km (labeled PREM-3) and 4 km (labeled PREM-4) and the {\it ak135} model (labeled {\it ak135}-3). The scale of the ratio of the PREM-3 to 
{\it ak135}-3 average densities is at most of order $\sim 10\%-15\%$. The impact of different densities on the energy loss parameters is small for all but the largest Earth emergence angles, as discussed in Appendix \ref{app:z2a}. Analogous to the impact of the water depth on the \tauon\ exit probability, we expect that the {\it ak135} model will yield \tauon\ exit probabilities within $\sim 10\%-15\%$ of the results presented here. Future versions of the \nupyprop\ code will include the {\it ak135} density model. One conclusion from these comparisons is that the local water depth for observations is the most important density effect at Earth emergence angles near the water-rock interface.
 
One of the largest modeling uncertainties comes from the photonuclear energy loss. Because the electromagnetic energy loss formulas are extrapolated well beyond kinematic regions that are measured, there are large variations in exit probabilities with different photonuclear interaction models. As shown in fig. \ref{fig:p-allm-bdhm}, for energies larger than $E_\nu\sim 10^9$ GeV, the choice of the {\texttt{allm}} or {\texttt{bdhm}} parameterization of the electromagnetic structure function $F_2$ can yield difference in the \tauon\ exit probabilities of $\sim 20\%-50\%$.

The impact of the photonuclear parameterizations on \tauon\ exit probabilities and on out-going energy distributions illustrate the usefulness of the \nupyprop\ code. Different codes have implementations that vary in many of the details, even if qualitatively, the same physics is included in the neutrino and charged lepton simulations. Small and large effects in modeling inputs can be quantified with this code. 
Future work will include further quantification of modeling uncertainties and their implications for instrument sensitivities, for example, of POEMMA  and EUSO-SPB2  sensitivities to target of opportunity astrophysical neutrino sources \cite{Venters:2019xwi,Reno:2021veo}.
 
\section{Summary}
\label{sec:summary}

We have introduced \nupyprop , a fast, modular and easy to use Monte Carlo package for $\nu_\tau \to \tau$ and $\nu_\mu\to \mu$ propagation inside the Earth.  It provides flexibility for the end user to change numerous free parameters. By not requiring any external dependencies for neutrino and charged lepton propagation and energy losses, the structure of the code is transparent and easy to read. 
We have shown selected results here.
The results from \nupyprop\ are in good agreement with other propagation codes. There are modeling systematic differences, so the use of multiple codes and approaches provide a way to quantify these systematic differences and their impact on neutrino detection for a given experimental configuration.

As a stand-alone code, \nupyprop\ produces lookup tables for the exit probabilities of \tauons\ and muons and their respective energy distributions. The lookup tables are for fixed incident neutrino energy and angle. Using interpolation routines with these lookup tables and standard Monte Carlo techniques, users can generate neutrino energy distributions based on theoretical neutrino flux predictions as inputs to detector simulations. 

The \nupyprop\ code has been developed as part of the 
\nuSpaceSim\ package \cite{Krizmanic:2021eyu}. The \nupyprop\ generated lookup tables are inputs to the extensive air shower simulations of \nuSpaceSim, key ingredients to the simulations of optical Cherenkov and geomagnetic radio signal modeling for the user-defined geometry and instrument response.
The modular features of \nupyprop\ will allow for quantitative assessments of theoretical uncertainties associated with neutrino and charged lepton propagation in the Earth in projected effective apertures and sensitivities for current and proposed instruments.

Installation of \nupyprop\ is described in \cref{sec:install}. The development of \nupyprop\ is on-going. Future additions will be tracking of lepton secondaries in neutrino interactions, for example, muons and muon neutrinos from $\nu_\tau\to \tau\to \nu_\tau+\mu+\bar{\nu}_\mu$, potentially important additions to the signals from $\nu_\tau$ and \tauons\ \cite{Beacom:2001xn,Dutta:2002zc}, as recently emphasized in ref. \cite{Arguelles:2022aum}. 

Active development and refinement of these codes like \nupyprop\ and \nuSpaceSim\ provide software tools to the community that works to design and realize the potential of new instruments to detect UHE cosmic neutrinos.
Measurements of these neutrinos will further our understanding of astrophysical sources and UHE neutrino interactions.

\vskip 0.1in

\noindent{\bf Acknowledgements}

\vskip 0.1in

This work is supported by NASA grants 80NSSC19K0626 at the University of Maryland, Baltimore County, 80NSSC19K0460 at the Colorado School of Mines, 80NSSC19K0484 at the University of Iowa, and 80NSSC19K0485 at the University of Utah, 80NSSC18K0464 at Lehman College, and  and RTOP 17-APRA17-0066 at NASA/GSFC and JPL.

\begin{appendix}

\section{Installation of \nupyprop}
\label{sec:install}
The \nupyprop\ code is open source and available at the
GitHub link \url{https://github.com/NuSpaceSim/nupyprop} and at \url{https://heasarc.gsfc.nasa.gov/docs/nuSpaceSim/} on the website of HEASARC.
The \nupyprop\ code can be installed with either pip or conda. 
With pip, use
\begin{lstlisting}
python3 -m pip install nupyprop
\end{lstlisting}
\noindent
to install.
With conda, we recommend installing \nupyprop\ into a conda environment. To install {\nupyprop} and activate the conda environment, use
\begin{lstlisting}
conda create -n nupyprop -c conda-forge -c nuspacesim nupyprop
conda activate nupyprop
\end{lstlisting}
\noindent
In this example, the name of the environment is “nupyprop”.

\section{How to Use \nupyprop}
\label{sec:usage}

Instructions are provided in the README file on GitHub. A plotting tutorial folder also exists in the GitHub repository, which provides a hands on tutorial on how to visualize the results and use different models of the simulation package from the output HDF files.
To list all the optional flags used in running the {\nupyprop} code, use
\begin{lstlisting}
nupyprop --help
\end{lstlisting}
\noindent 
which will show the following, 
\begin{enumerate}
\item
   \texttt{-e} or \texttt{--energy}: incoming neutrino energy in $\log_{10}$(GeV). Works for single energy or multiple energies. For multiple energies, separate energies with commas, e.g., \texttt{-e 6,7,8,9,10,11}. Default energies are $10^6$ to $10^{11}$ GeV, in steps of quarter decades. 
\item   
    \texttt{-a} or \texttt{--angle}: Earth emergence angles in degrees. Works for single angle or multiple angles. For multiple angles, separate angles with commas, e.g., \texttt{-a 0.1,0.2,1,3,5,7,10}. Default angles are 1 to 42 degrees, in steps of 1 degree.
    \item
    \texttt{-i} or \texttt{--idepth}: depth of ice/water in km, e.g, \texttt{-i 3} for depth of 3 km. Default value is 4 km. Other options range between 0 km and 10 km in integer units.
\item    
    \texttt{-cl} or \texttt{--charged\_lepton}: flavor of charged lepton used to propagate. Can be either muon or tau. Default is \texttt{-cl tau}.
\item    
    \texttt{-n} or \texttt{--nu\_type}: type of neutral lepton, either neutrino or anti-neutrino. Default is \texttt{-n neutrino}. For energies above $E_\nu\sim 10^7$, neutrino and anti-neutrino results are nearly identical.
\item    
    \texttt{-t} or \texttt{--energy\_loss}: energy loss type for lepton - can be stochastic or continuous. Default is \texttt{-t stochastic}.
\item
    \texttt{-x} or \texttt{--xc\_model}: neutrino/anti-neutrino cross-section model used. Can be from the pre-defined set of models (ct18nlo,nct15,ctw,allm,bdhm) or custom (see~\cref{app:custom}). Default is \texttt{-x ct18nlo}.
\item
    \texttt{-p} or \texttt{--pn\_model}: lepton photonuclear energy loss model, either bdhm or allm. Default is \texttt{-p allm}.
\item
    \texttt{-el} or \texttt{--energy\_lepton}: option to print each exiting charged lepton's final energy in output HDF file. Default is \texttt{-el no}.
\item
    \texttt{-f} or \texttt{--fac\_nu}: rescaling factor to approximate BSM modified neutrino cross sections. This modifies the cross section only, not the CDF for the outgoing lepton energy distribution. Default is \texttt{-f 1}.
\item
    \texttt{-s} or \texttt{--stats}: statistics (number of incident neutrinos of a given energy). Default is $10^7$ neutrinos: \texttt{-s 1e7}.
\item
    \texttt{-htc} or \texttt{--htc\_mode}: High throughput computing (HTC) mode. The code can be made to run in HTC mode on a local cluster by the user, as {\nupyprop} is designed in a way that it can be easily run in a parallel computation environment. If set to yes, {\nupyprop} will only create separate data files for final energy, average polarization, CDFs, and exit probabilities of exiting charged leptons, and not an HDF file. These data files once created for the desired statistics can then be post-processed by using the function process\_htc\_out() residing in data.py code. Default is {\texttt{-htc no}}. 
\end{enumerate}
\noindent
An example command for running tau neutrino and tau propagation for neutrino energy $E_\nu=10^7$ GeV, incident with a 10 degree Earth emergence angle for $10^6$ neutrinos injected with stochastic energy loss, with all other parameters as defaults,
\begin{lstlisting}
nupyprop -e 7 -a 10 -t stochastic -s 1e6
\end{lstlisting}
\noindent
which yields output in a file name as, 
\texttt{output\_nu\_tau\_4km\_ct18nlo\_allm\_stochastic\_1e6.h5},
in the directory in which the 
\nupyprop\ run command is executed. 

\section{Customization of input lookup tables}
\label{app:custom}
{\nupyprop} uses lookup tables or ecsv files (cross-section, CDF, and for photonuclear interactions, beta, i.e. energy loss parameter, ecsv file) to call models for neutrino and electromagnetic interactions. The models included in the lookup table for neutrino interaction are, \texttt{allm}, \texttt{bdhm}, \texttt{ct18nlo}, and \texttt{nct15} and for electromagnetic interactions, they are \texttt{allm} and \texttt{bb}. If the user wants to use a custom model, then {\nupyprop} will look for an ecsv file, specific to the custom model's name, in src/nupyprop/models directory. To create an ecsv file, user has to create a function for their custom model in models.py code. 
There are examples in the code to explain the structure of the custom model. One example shows a neutrino interaction model, \texttt{ctw} model, and the another shows photonuclear interaction models, \texttt{bdhm} and \texttt{ckmt}. The
user can execute models.py code with added custom subroutines to create the respective ecsv file for the custom model.
Once the ecsv file is created for the custom model, run the {\nupyprop} code on the command line with the custom model's name in it. 

\section{Supplemental material}
\label{app:supplement}

\subsection{Neutrino cross section parameterizations}
Connolly, Thorne and Waters \cite{Connolly:2011vc} parameterized their neutrino cross section in terms of $\epsilon = \log_{10}(E_\nu/{\rm GeV})$, with the cross section written as
\begin{equation}
 \label{eq:xc_nu}
 \log_{10}[\sigma(\epsilon)/{\rm cm}^2] = C_1+C_2\cdot \ln (\epsilon-C_0)
 +C_3\cdot \ln^2(\epsilon-C_0)+C_4/\ln(\epsilon-C_0)\ ,
\end{equation}
for $E_\nu=10^4-10^{12}$ GeV. For reference, we provide numerical best fits for these parameters listed in~\cref{tab:neutrinoxc} for the CC and NC cross sections for $\nu$ and $\bar{\nu}$ interactions with isoscalar nucleons. The \texttt{allm} neutrino cross section is normalized such that it is equal to the \texttt{ct18nlo} cross section at $E_\nu=10^7$ GeV. The \texttt{allm} and \texttt{bdhm} should be used only for $E_\nu\gtrsim 10^6$ GeV.

\begin{table}[htbp!]
\label{table:xc_nu_params}
    \centering
    \begin{tabular}{|c|c|c|c|c|c|}
    \hline
    \texttt{allm} \cite{Abramowicz:1997ms} & $\mathbf{C_0}$ & $\mathbf{C_1}$ & $\mathbf{C_2}$ & $\mathbf{C_3}$ & $\mathbf{C_4}$\\ \hline
    $\nu$ CC &   2.558 &   -35.239 &   1.230 &   0.224 & -0.002\\ \hline
    $\nu$ NC &  2.889  &  -35.284  &  1.029  & 0.291  & 0.001 \\ \hline
    $\bar{\nu}$ CC & 2.558 &   -35.239 &   1.230 &   0.224 & -0.002 \\ \hline
    $\bar{\nu}$ NC &  2.889  &  -35.284  &  1.029  & 0.291  & 0.001 \\
    \hline
    \texttt{bdhm} \cite{Block:2014kza} & & & & &\\ \hline
     $\nu$ CC &   1.062 &   -41.540 &   5.912 &   -0.796 & 1.362\\ \hline
     $\nu$ NC &  0.982  &  -43.082  &  6.652  & -0.929  & 1.748 \\ \hline
     $\bar{\nu}$ CC & 1.062 &   -41.540 &   5.912 &   -0.796 & 1.362 \\ \hline
     $\bar{\nu}$ NC &  0.982  &  -43.082  &  6.652  & -0.929  & 1.748 \\ \hline
    \texttt{ct18nlo} \cite{Hou:2019efy} & & & & &\\ \hline
     $\nu$ CC &   -1.800 &   -19.478 &   -5.812 &   1.417 & -15.871\\ \hline
     $\nu$ NC &  -2.247  &  -12.089  &  -8.703  & 1.789  & -23.426 \\ \hline
     $\bar{\nu}$ CC & 2.663  &  -34.826  &  0.951  & 0.331  & -$3.275 \times 10^{-4}$ \\ \hline
     $\bar{\nu}$ NC &  2.656  &  -35.285  & 0.965   & 0.339 & -$2.416 \times 10^{-4}$ \\ \hline
    \texttt{ctw} \cite{Connolly:2011vc} & & & & &\\ \hline
     $\nu$ CC &   -1.826 &   -17.310 &   -6.406 &   1.431 & -17.910\\ \hline
     $\nu$ NC &  -1.826  &  -17.310  &  -6.448  & 1.431  & 18.610 \\ \hline
     $\bar{\nu}$ CC & 2.443  &  -35.104  &  1.167  & 0.260  & $4.031 \times 10^{-4}$ \\ \hline
     $\bar{\nu}$ NC &  2.536  &  -35.453  & 1.162   & 0.267 & $3.045 \times 10^{-4}$ \\ \hline
    \texttt{nct15} \cite{Kovarik:2015cma} & & & & &\\ \hline
     $\nu$ CC &   -2.018 &   -2.042 &   -14.376 &   2.813 & -27.906\\ \hline
     $\nu$ NC &  -4.423  &  86.933  &  -47.161  & 6.840  & -111.423 \\ \hline
     $\bar{\nu}$ CC & 2.158  &  -35.295  &  1.061  & 0.355  & -$5.487 \times 10^{-4}$ \\ \hline
     $\bar{\nu}$ NC &  2.376  &  -35.526  & 1.021   & 0.365  & -$4.407 \times 10^{-4}$ \\ \hline
    \end{tabular}
    \caption{Constants for the parameterization in  \cref{eq:xc_nu} of the neutrino and antineutrino cross sections for charged-current (CC) and neutral-current (NC) interactions with isoscalar nucleons.  The parameterizations are valid for $10^4$ GeV$<E_\nu<10^{12}$ GeV for the PDF based cross sections. The cross sections labeled \texttt{allm} and \texttt{bdhm} can be used for $E_\nu\gtrsim 10^6$ GeV, where $\sigma^{\nu N}\simeq \sigma^{\bar\nu N}$ and valence quark contributions are negligible. }
    \label{tab:neutrinoxc}
\end{table}

\subsection{Density dependent corrections for bremsstrahlung and pair production electromagnetic energy loss}
\label{app:z2a}

The density model of the Earth does not dictate its chemical composition as a function of radial distance from the center. Here, we approximate the core as being composed of iron ($Z=26$, $A=56$), and apart from a surface water layer, we approximate the crust as primarily rock ($Z=11$, $A=22$). For each of the energy loss mechanisms, we have evaluated the energy loss parameter $\beta^\ell$ as a function of $Z$ and $A$ to determine approximate interpolations between the electromagnetic energy loss parameters and interaction depths for each of the contributions: ionization, bremsstrahlung, pair production and photonuclear interactions.

For both bremsstrahlung and pair production, there is nominally a $Z^2$ dependence to the differential cross section and a factor of $1/A$ for $\beta^\ell_{\rm pair}$ and the interaction depth. Numerical evaluation of $\beta^\ell_{\rm pair}$ in the range from rock to iron, for which there is additional $Z$ dependence (see, e.g., ref. \cite{Lohmann:1985qg}) for muons and taus, dictates a modification to 
$
    \beta^\ell_{\rm brem} \sim {Z^{1.86}}/{A}
$
and $\beta^\ell_{\rm pair} \sim {Z^{1.87}}/{A}$.
The code has lookup tables for electromagnetic energy loss via pair production for water and rock. For $N_A=6.022\times 10^{23}$, the iron energy loss parameter and cross sections used here are
\begin{eqnarray}
    \beta^\ell_i({\rm Fe}) &\simeq& \Biggl(\frac{26}{11}\Biggr)^{1.87}
    \frac{22}{56}\cdot \beta^\ell_i({\rm rock})\simeq 1.97
    \beta^\ell_i({\rm rock})\,, \quad\quad i={\rm pair,brem}\\ 
    \frac{N_A\sigma_i({\rm Fe})}{56}&\simeq & 1.97\, \frac{N_A\sigma_i({\rm rock})}{22}\,,\quad\quad i={\rm pair,brem}\,.
\end{eqnarray}

Electromagnetic energy loss per nucleon via photonuclear interactions is largely insensitive to $Z$ and $A$. We find that due to nuclear shadowing, electromagnetic energy loss in iron is about 10\% lower than energy loss in rock and similarly for the cross section:
\begin{eqnarray}
    \beta^\ell_{\rm nuc}({\rm Fe})&\simeq& \Biggl(\frac{56}{22}\Biggr)^{-0.1}
    \beta^\ell_{\rm nuc}({\rm rock})\simeq 0.91\, \beta^\ell_{\rm nuc}({\rm rock})\\
     \frac{N_A\sigma_{\rm nuc}({\rm Fe})}{56}&\simeq & 0.91\, \frac{N_A\sigma_{\rm nuc}({\rm rock})}{22}
\end{eqnarray}

Within the Earth, for $\rho>\rho_{\rm Fe}= 7.87$ g/cm$^3$ ($\beta>56.9^\circ$), we still use the beta and cross-section values of the iron by using the scaling from rock to iron as shown in the above equations. 
For densities between that of iron and of
rock $\rho_{\rm rock}=2.60$ g/cm$^3$ (the minimum density in the PREM model apart from water), 
we approximate by density fraction according to 
\begin{eqnarray}
f_{\rm rock}&=& \frac{(\rho_{\rm Fe}-\rho)}{(\rho_{\rm Fe} -\rho_{\rm rock})}\\
    \beta_i&=& f_{\rm rock}\beta_i^{\rm rock}
    + (1-f_{\rm rock})\beta_i^{\rm Fe}\label{eq:rockfrac}\\
 \beta_i   &=& \beta_i^{\rm rock}(1.97 - 0.97\, f_{\rm rock})\quad\quad i={\rm brem,pair}\\
 \beta_{\rm nuc} &=&  \beta_{\rm nuc}^{\rm rock}(0.91+0.09\, f_{\rm rock})\,,
\end{eqnarray}
and similarly for the cross sections (inverse interaction lengths).
Our numerical results show that these density corrections to \tauon\ energy loss do not change the exit probabilities or energy distributions by more than $\sim 5\%$ for all but the largest Earth emergence angles, where statistical uncertainties in our evaluation make it difficult to determine the effect.

\subsection{Tau neutrino energy distributions from \tauon\ decays}
\label{subsec:taudecay}

The \tauon\ decay distribution to $\nu_\tau$ is required to incorporate regeneration. In this appendix, we discuss modeling the energy distribution of the decay neutrino and show that, to a good approximation, the $\nu_\tau$ distribution follows its distribution in the purely leptonic channel.

For the decay distributions for each decay channel, we define $y_\nu=E_\nu/E_\tau$ in terms of the tau neutrino and \tauon\ energies in the lab frame.
We begin with the leptonic decays of the $\tau^-$.
The energy distribution of the tau neutrino in purely leptonic decays in the relativistic limit is \cite{Barr:1989ru,Gaisser:2016uoy},
\begin{eqnarray}
    \frac{1}{\Gamma}\frac{d\Gamma_\ell}{dy_\nu} &=& B_\ell
 \  \Bigl[ g_0^\ell(y_\nu)+{\cal P}_{z} g_1^\ell(y_\nu)\Bigr]\theta(y_\nu^{\rm max}-y_\nu)
\end{eqnarray}
in terms of the branching fraction $B_\ell=0.18$ for $\ell = e$ and $\ell=\mu$, and functions $g_0^\ell (y_\nu)$ and $g_1^\ell (y_\nu)$ in Table \ref{table:taudecays}. 
Here ${\cal P}_{z}$ indicates the projection of the tau spin in its rest frame along the direction of motion of the tau in the lab frame. For left-handed $\tau^-$, ${\cal P}_{z}=-1$. The $\theta$-function enforces the upper bound on the ratio of $E_\nu$ to $E_\tau$, which is unity in the massless $\ell=e$ and $\ell=\mu$ limits assumed here. The same form of the $\bar\nu_\tau$ energy distribution in $\tau^+$ decays follows for the antiparticle decay, $d\Gamma_{\tau^+}\sim [g_0^\ell-{\cal P}_{z} g_1^\ell]$, with ${\cal P}_{z}=+1$ for right-handed $\tau^+$. This equivalence applies to the semi-leptonic decays as well.

The decay $\tau^-\to \nu_\tau \pi^-$ is straightforward to describe with the two-body decay functions $g_0^\pi$ and $g_1^\pi$ in terms of $y_\nu$ and $r_\pi\equiv m_\pi^2/m_\tau^2$. The maximum of $E_\nu$ is $E_\tau (1-r_\pi)$.

The decays  $\tau^-\to \nu_\tau \rho^-$ follows similarly to $\tau^-\to\nu_\tau \pi^-$, but with the addition of smearing by a p-wave Breit-Wigner factor that depends on $Q^2$.
Following Kuhn and Santamaria \cite{Kuhn:1990ad} extended to include the \tauon\ polarization, the decay $\tau^-\to \nu_\tau\rho^-$ can be approximated by,
\begin{equation}
    \frac{d{\Gamma}_\rho}{dy_\nu} = N_\rho\int_{r_{\rm min}}^1 dr_Q\, \rho(Q^2) (1-r_Q)^2(1+2r_Q)[f_0^Q+ {\cal P}_{z}f_1^Q]\Biggl(1-\frac{r_{\rm min}}{r_Q}\Biggr)^{3/2}\theta(y_{\rm max}-y_\nu)
    \, ,
    \label{eq:taudecaysmeared}
\end{equation}
including $N_\rho$, the overall normalization factor that accounts for the branching fraction $B_\rho$ and where $r_Q=Q^2/m_\tau^2$ and $r_{\rm min}=Q^2_{\rm min}/m_\tau^2$ with $Q^2_{\rm min}=(2m_\pi)^2$.
In its simplest form, the $Q^2$ dependent function $\rho(Q^2)$ can be written \cite{Kuhn:1990ad,Shekhovtsova:2017nof}
\begin{eqnarray}
\rho(Q^2) &=& |BW_\rho(Q^2)|^2\\
BW_\rho(Q^2) &=& \frac{m_\rho^2}{m_\rho^2-Q^2-im_\rho\Gamma_{\rm tot,\rho}(Q^2)}\\
\Gamma_{\rm tot,\rho}(Q^2)&=& \Gamma_{\rho,0}\frac{Q^2}{m_\rho^2}
 \Biggl(\frac{1-Q_{\rm min}^2/Q^2}{1-Q_{\rm min}^2/m_\rho^2}\Biggr)^{3/2}\ ,
\end{eqnarray}
where $Q^2_{\rm min}=(2m_\pi)^2$.
In fact, $BW_\rho(Q^2)\to B(Q^2)$ in eq. (\ref{eq:taudecaysmeared}), where $B(Q^2)$ is a combination of three Breit-Wigner functions 
according to 
\begin{eqnarray}
    B(Q^2) =
\frac{1}{1+\beta+\delta}[BW_{\rho''}+\beta BW_{\rho'} +\delta BW_\rho]\,,
\end{eqnarray}
used in TAUOLA v2.4 \cite{Jadach:1993hs} and  also used here.
The masses and widths appear in Table \ref{tab:masswidth}.
We also follow ref. \cite{Jadach:1993hs} to include $\tau^-\to a_1^-\nu_\tau $ with 
\cite{Kuhn:1990ad} 
\begin{equation}
\label{eq:a1}
\frac{d\Gamma}{dy_\nu}= N_{a1}\int_{r_{\rm min}}^1 \, dr_Q\,  (1-r_Q)^2(1+2r_Q) [f_0^Q + {\cal P}_{z}f_1^Q]  |BW_{a1}(rm_\tau^2)|^2 \frac{g(rm_\tau^2)}{rm_\tau^2}\theta(y_{\rm max}-y_\nu)\,,
\end{equation}
where $g(Q^2)$ is defined in eqn. (3.16) of ref. \cite{Kuhn:1990ad}. In the Breit-Wigner function, 
\begin{equation}
\label{eq:ga1}
\Gamma(Q^2)=\Gamma_{a1}\frac{g(Q^2)}{g(m_{a1}^2)}\,.
\end{equation}
For the $\tau^-\to \nu_\tau 4\pi$ decay, we use eqs. (\ref{eq:a1}) and (\ref{eq:ga1}) substituting mass, width and $Q_{\rm min}$ according to Table \ref{tab:masswidth}.

\begin{figure}
    \centering
    \includegraphics[width=0.6\textwidth]{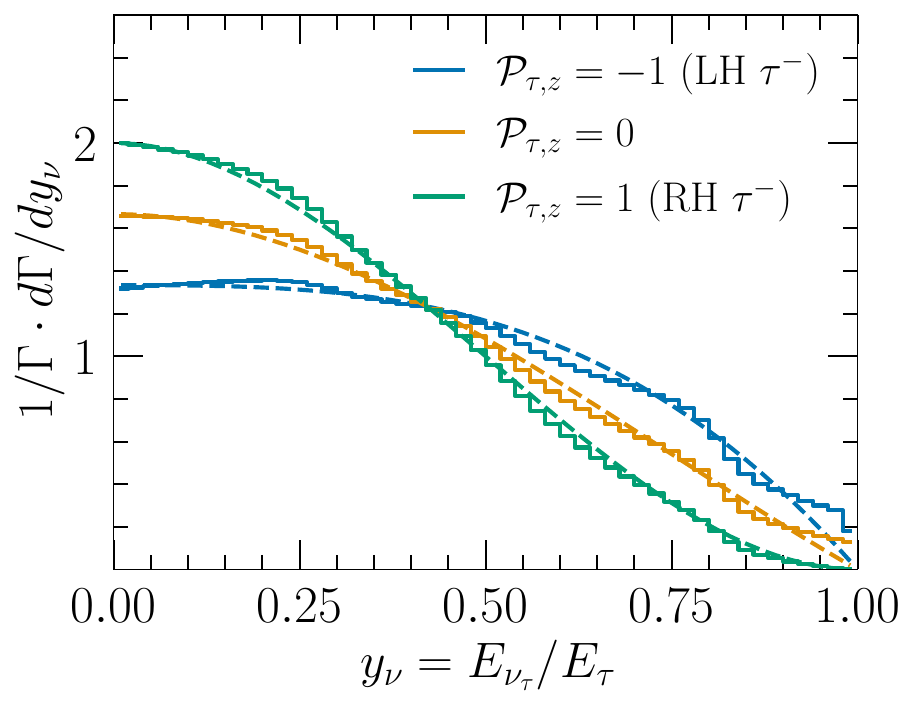}
    \caption{The energy distribution $y=E_\nu/E_\tau$ for three tau polarizations.
    The histograms show approximate energy distributions from the sum of leptonic and semileptonic modes discussed in the text, and the dashed curves are the normalized tau neutrino energy distribution from the purely leptonic decay mode.}
    \label{fig:taudecay}
    \vspace{-10pt}
\end{figure}

Figure \ref{fig:taudecay} shows the distribution $1/\Gamma\, d\Gamma/dy_\nu$ including Breit-Wigner smearing for $\Delta y=0.02$ bins for left-handed \tauons\ (blue histogram). For reference, also shown are the distributions for right-handed \tauons\ (green histogram) and unpolarized \tauons\  (orange histogram). Overlaid are the respective $y_\nu$ distributions from purely leptonic decays (dashed), normalized to unity. Since we require only the inclusive tau neutrino energy distribution, given how well the leptonic distribution matches the distribution in $y_\nu$ from the sum over all of the decay channels as approximated here, we use $g_0^\ell + {\cal P}_{z} g_1^{\ell}$ to describe the $y_\nu=E_\nu/E_\tau$ distributions from LH $\tau^-$ decays. 
The cumulative distribution function of the tau neutrino energy fraction is \begin{eqnarray}
\label{eq:cdfnu}
    {\rm CDF}(y_\nu) &=& \int_0^{y_\nu} dy' \ \Bigl[g_0^\ell (y')+{\cal P}_{z} g_1^\ell (y')\Bigr]\nonumber \\
    &=& \frac{5}{3}y_\nu-y_\nu^3+\frac{y_\nu^4}{3}+{\cal P}_{\tau ,z}\Biggl(\frac{y_\nu}{3}-y_\nu^3+\frac{2}{3}y_\nu^4\Biggr)\ .
\end{eqnarray}
The decay distribution for the tau antineutrino from $\tau^+$ decay has the same CDF. 
Eq. (\ref{eq:cdfnu}) simplifies the evaluation of the neutrino energy when the tau decays in \nupyprop.

\begin{table}
\begin{tabular}{|l|c|c|c|c|}
\hline
Process & $B_i$ & $g_0^i$ & $g_1^i$  & $y_{max}$ \\ \hline
$\tau\rightarrow \nu_\tau \mu \nu_\mu$ & 0.18  &  ${5/ 3} - 
3y^2+{4}y^3/3$
& ${1/ 3} - 3y^2+{8}y^3/3 $  & 1 \\ \hline
$\tau\rightarrow \nu_\tau e\nu_e$ & 0.18  &  ${5/ 3} - 
3y^2+{4}y^3/3$
& ${1/ 3} - 3y^2+{8}y^3/3 $ & 1 \\ \hline
$\tau\rightarrow \nu_\tau \pi$ & 0.12 & ${ (1-r_\pi)^{-1}}$ 
& $-{(2y -1+r_\pi)/ (1-r_\pi)^2}$ &$ (1-r_\pi)$  \\ \hline
$\tau\rightarrow \nu_\tau \rho$ & 0.26 & $f_0^Q={( 1-r_Q)^{-1}}$ 
& $f_1^Q=-(2y-1+r_Q)(1-2r_Q)$ & $ (1-r_Q)$  \\  
& & & $\times[ (1-r_Q)^{2}
(1+2r_Q)]^{-1}$ & \\ \hline
$\tau \rightarrow  \nu_\tau a_1$ & 0.19 & $f_0^Q$ 
& $f_1^Q$ & $(1-r_Q)$ \\ \hline
$\tau\to 4\pi$& 0.07 & $f_0^Q$ & $f_1^Q$ & $(1-r_Q)$ \\ \hline
\end{tabular}
\caption{For $i=e,\mu,\pi,\rho,a_1$ and $4\pi$, the functions $g_0^i$ and $g_1^i$ \cite{Pasquali:1998xf,Bhattacharya:2016jce}
and the branching fractions in the tau neutrino energy distribution 
from relativistic \tauon\ decays, in terms of $y_\nu=y=E_\nu/E_\tau$ and $r_\pi=m_\pi^2/m_\tau^2$
and $r_Q=Q^2/m_\tau^2$ for the remaining decays that are smeared with a Breit-Wigner factor. }
\label{table:taudecays}
\end{table}

\begin{table}[htp]
\begin{center}
\begin{tabular}{|cc|c|c|c|}
\hline
Particle & &  $m$ [GeV] & $\Gamma$ [GeV] & $Q_{\rm min}$\\
\hline
\hline
$\pi$ & &  0.14 & - &-\\
\hline
$\rho$ & &  0.773  & 0.145 & $2m_\pi$\\
$\rho'$ & &  1.37 & 0.510 &\\
$\rho''$ & &  1.75 & 0.12 & \\
\hline
$a_1$ &&  1.26 & 0.25 & $3m_\pi$\\
\hline
$4\pi$ &&  1.50 & 0.25 & $4m_\pi$\\
\hline
\end{tabular}
\end{center}
\caption{Parameters for semi-leptonic decays of $\tau^-$ \cite{Kuhn:1990ad}.}
\label{tab:masswidth}
\end{table}

\subsection{Polarization of \tauons}
\label{subsec:polarization}

A detailed study of the polarization effects on the exiting {\tauons} is described in ref. \cite{Arguelles:2022bma}. Polarization is only considered for {\tauons} and not muons because the long lifetime of the muon leads to large energy losses before its decay, producing a low energy regenerated $\nu_\mu$, which is not of interest to the neutrino telescope experiments.
To a very good approximation, the average polarization is independent of final energy of exiting {\tauons} \cite{Arguelles:2022bma}. The average \tauon\ polarization is a function of the Earth emergence angle and initial $\nu_\tau$ energy. 

\begin{figure*}[ht!]
    \centering
    \includegraphics[width=0.7\textwidth]{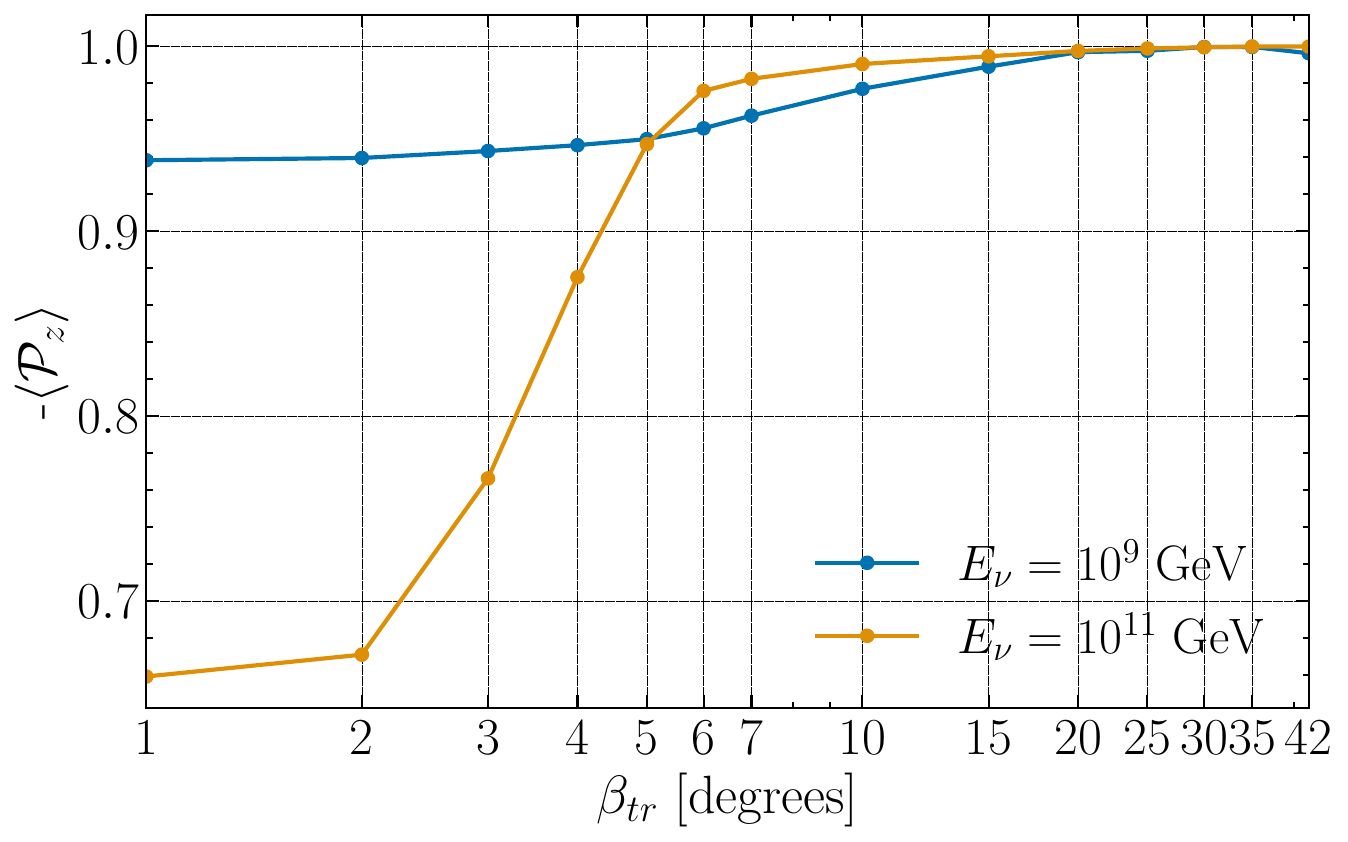}
    \caption{The negative of the average polarization of exiting {\tauons} $(-\langle {\cal P}_z\rangle)$ as a function of Earth emergence angles ($\beta_{tr}$)using \nupyprop, for two different initial $\nu_\tau$ energies \cite{Arguelles:2022bma}. Left-handed \tauons\ have
    $\langle {\cal P}_z\rangle=-1$.}
    \label{fig:avg_polarization}
\end{figure*}

In~\cref{fig:avg_polarization}, we show the average polarization about z-axis, $\langle {\cal P}_z\rangle$, of the exiting {\tauons} as a function of the Earth emergence angles, for two different initial $\nu_\tau$ energies, $10^9$ GeV and $10^{11}$ GeV. This can be used as an input to the decay of the \tauon\ that produces an EAS.
Below $\beta_{tr}\lesssim 4^\circ$, regeneration is negligible for all energies. For $E_\nu=10^9$~GeV, regeneration occurs for $\beta\gtrsim 10^\circ$, while for $E_\nu=10^{11}$~GeV, regeneration occurs for $\beta\gtrsim 4^\circ$ (see~\cref{fig:p-regen}).
For the angles where there is no regeneration, the exiting {\tauons} are created from the initial $\nu_\tau$ and for high energy \tauons\, their long trajectories prior to exiting the Earth cause them to be somewhat depolarized. For lower energies, the \tauons\ that exit the Earth are produced close to the surface so do not have many depolarizing interactions.
For angles where there is regeneration, the {\tauons} that exit the Earth are produced from regenerated $\nu_\tau$. Each \tauon\ produced in CC interactions has the polarization reset to left-handed (${\cal P}_z=-1$), so \tauons\ from regeneration are mostly polarized. 

\begin{figure*}[ht!]
    \centering
    \includegraphics[width=0.8\textwidth]{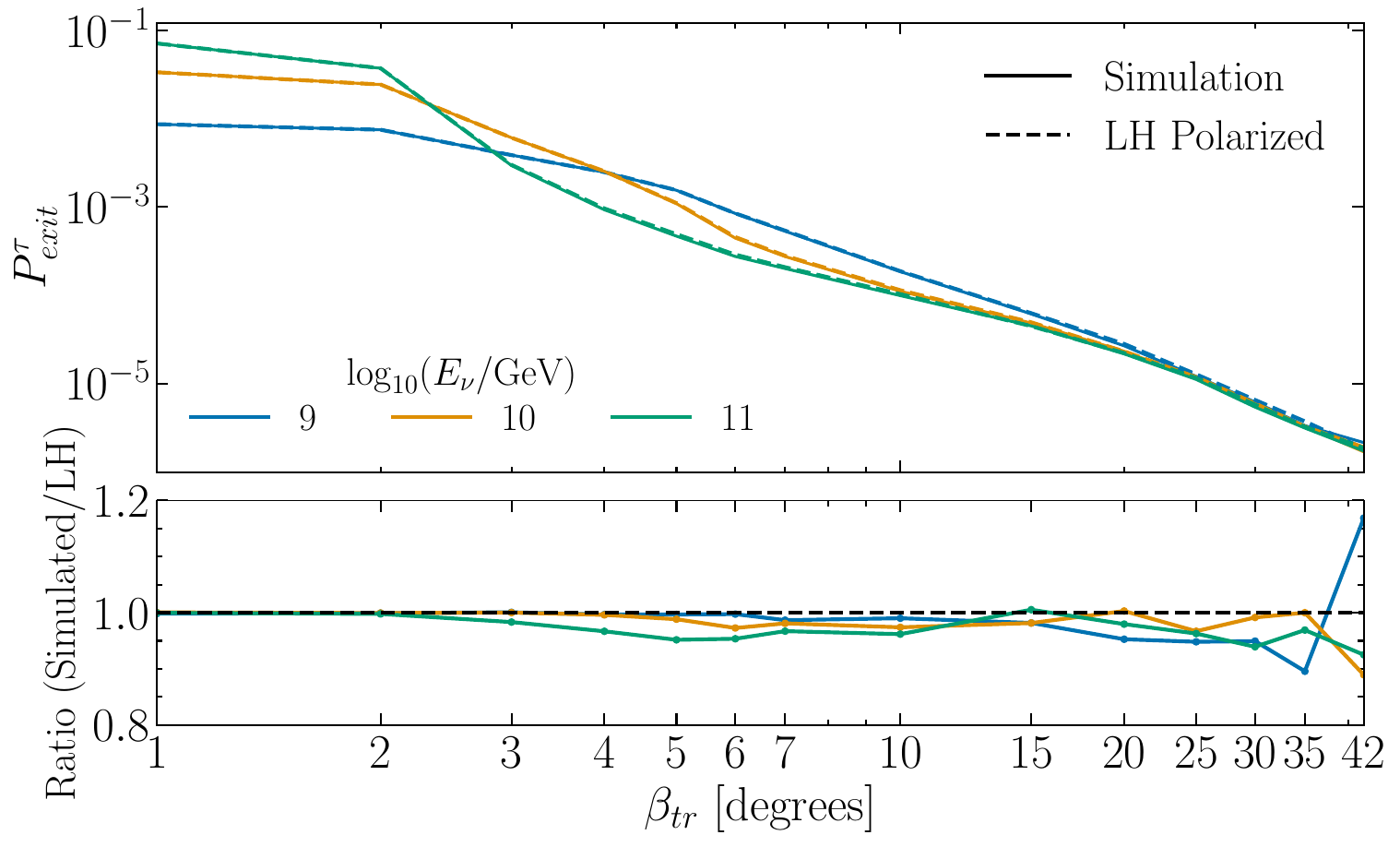}
    \caption{Exit probability of {\tauons} as a function of Earth emergence angles using \nupyprop, for three different initial $\nu_\tau$ energies. It shows a comparison when we consider LH polarization and simulated depolarization for EM interactions of the taus \cite{Arguelles:2022bma} in \tauon\ decays in the evaluation of regeneration contributions. }
    \label{fig:pexit_polarization}
\end{figure*}

\Cref{fig:pexit_polarization} shows the exit probability for {\tauons} as a function of the Earth emergence angle. It shows a comparison between using LH polarization ($\langle {\cal P}_z\rangle =-1$) and using simulated depolarization from EM interactions in the {\tauons} decay distributions for regeneration. It shows us that the depolarization has a $\sim 5\%$ effect for smaller angles and $\sim 10\%$ effect for larger angles. Overall, the depolarization of \tauons\  from EM interactions has a small impact on the exit probability of the {\tauons}.  

\end{appendix}

\vfil\eject
\bibliographystyle{JHEP}
\bibliography{references}
\end{document}